\documentclass[lettersize,journal]{IEEEtran} 
\usepackage{amsmath}
\usepackage{amsthm}
\usepackage{multirow}
\usepackage{color,cite}
\usepackage{subfigure}
\usepackage{mathrsfs} 
\usepackage{amssymb}
\usepackage{bm}
\usepackage{graphicx}
\usepackage{amsfonts}
\usepackage{algorithm}
\usepackage{algorithmic}
\usepackage{makecell}
\usepackage{etoolbox}

\DeclareMathOperator*{\argmin}{arg\,min}

\usepackage{algorithm}
\usepackage{algorithmic}
\usepackage{makecell}
\usepackage{etoolbox}
\usepackage{geometry}
\usepackage{stfloats}

\usepackage{bm}
\usepackage{setspace}

\input epsf

\begin{document}
\title{{Cross-Domain Dual-Functional OFDM Waveform \\ Design for Accurate Sensing/Positioning}}

\author{ 
 {Fan~Zhang,~\IEEEmembership{Student Member,~IEEE}, 
Tianqi~Mao,~\IEEEmembership{Member,~IEEE}, Ruiqi~Liu,~\IEEEmembership{Member,~IEEE},\\
Zhu~Han,~\IEEEmembership{Fellow,~IEEE}, 
Sheng~Chen,~\IEEEmembership{Life~Fellow,~IEEE},
and Zhaocheng~Wang,~\IEEEmembership{Fellow,~IEEE}}
\thanks{This work was supported in part by the National Key R$\&$D Program of China under Grant 2021YFA0716603,
in part by National Natural Science Foundation of China under Grant No. 62088101,
and in part by Young Elite Scientists Sponsorship Program by CAST under Grant 2022QNRC001. 
Part of this work has been submitted to IEEE WCNC 2024 \cite{con}. 
\emph{(Corresponding authors: Zhaocheng Wang, Tianqi Mao.)}} 
\thanks{F. Zhang and Z. Wang are with Beijing National Research Center for Information Science and Technology, Department of Electronic Engineering, Tsinghua University, Beijing 100084, China, and Z. Wang is also with Tsinghua Shenzhen International Graduate School, Shenzhen 518055, China (e-mails: zf22@mails.tsinghua.edu.cn, zcwang@tsinghua.edu.cn).} %
\thanks{Tianqi Mao is with the State Key Laboratory of CNS/ATM, Beijing Institute of Technology, Beijing 100081, China, and also with the MIIT Key Laboratory of Complex-field Intelligent Sensing, Beijing Institute of Technology, Beijing 100081, China (e-mail: maotq@bit.edu.cn).}
\thanks{R. Liu is with the Wireless and Computing Research Institute, ZTE Corporation, Beijing 100029, China, and also with the State Key Laboratory of Mobile Network and Mobile Multimedia Technology, Shenzhen 518055, China (e-mail: richie.leo@zte.com.cn).} %
\thanks{Z. Han is with the Department of Electrical and Computer Engineering, University of Houston, Houston, TX 77004 USA, and also with the Department of Computer Science and Engineering, Kyung Hee University, Seoul 446-701, South Korea (e-mail: hanzhu22@gmail.com).} %
\thanks{S. Chen is with the School of Electronics and Computer Science, University of Southampton, Southampton SO17 1BJ, U.K. (e-mail: sqc@ecs.soton.ac.uk).} %
\vspace*{-7mm}
}

\maketitle
 
\begin{abstract}
Orthogonal frequency division multiplexing (OFDM) has been widely recognized as the representative waveform for 5G wireless networks, which can directly support sensing/positioning with existing infrastructure. 
To guarantee superior sensing/positioning accuracy while supporting high-speed communications simultaneously,
the dual functions tend to be assigned with different resource elements (REs) due to their diverse design requirements. This motivates optimization of resource allocation/waveform design across time, frequency, power and delay-Doppler domains. Therefore, this article proposes two cross-domain waveform optimization strategies for effective convergence of OFDM-based communications and sensing/positioning, following communication- and sensing-centric criteria, respectively. For the communication-centric design, to maximize the achievable data rate, a fraction of REs are optimally allocated for communications according to prior knowledge of the communication channel. 
The remaining REs are then employed for sensing/positioning, where the sidelobe level and peak-to-average power ratio are suppressed by optimizing its power-frequency and phase-frequency characteristics for sensing performance improvement. 
For the sensing-centric design, a `locally' perfect auto-correlation property is ensured for accurate sensing and positioning by adjusting the unit cells of the ambiguity function within its region of interest (RoI). Afterwards, the irrelevant cells beyond RoI, which can readily determine the sensing power allocation, are optimized with the communication power allocation to enhance the achievable data rate. Numerical results demonstrate the superiority of the proposed 
waveform designs.
\end{abstract}


\vspace{-4mm}
\section{Introduction}\label{S1}
 
With the commercialization of the fifth-generation (5G) networks, early explorations of the game-changing sixth-generation (6G) concept have been initiated by a collection of countries, being envisioned to support unprecedented ubiquitous sensing/positioning tasks aside from communications \cite{6G_cst_23,Saad_Network_20,Chowdhury_ojcoms_20,itu2023vision}. Promoted by the ever-progressing digital signal processing techniques, the transceiver structure for radar sensing and data transmission has become increasingly correlated \cite{Sturm_procieee_11,H_JSAC_22,X_JSAC_22}. This 
eventually enables direct target sensing/positioning with the existing infrastructure of wireless cellular networks, yielding the dual-functional radar and communication (DFRC) philosophy \cite{Liu_tcom_20,Andrew_survey_22,Cui_network_21}. Such cutting-edge technology can realize both functions simultaneously with identical hardware resources and efficient bandwidth/energy usage, which supports a plethora of emerging applications, e.g., autonomous driving, smart home, extended reality (XR), airborne reconnaissance/monitoring, etc.  \cite{Liu_jsac_22,ISACadd1,ISACadd2}. Thanks to these merits, DFRC has attracted extensive attentions as a promising enabling technology for 6G wireless networks and is recognized as one of six independent usage scenarios of 6G \cite{COMMAG_6G_early, tnsm_early, E_tcom_23}.

Existing literature has reached a consensus that the broad-sense concept of DFRC can be classified into three categories: co-existence, cooperation and co-design  \cite{Chiriyath_tccn_17,Feng_chinacom_20}. Co-existing radar and communication subsystems are mutually treated as independent and adverse interferers, whilst the cooperation counterpart can mitigate the interference through information exchange. However, these two approaches do not fully integrate sensing and communication subsystems, which induces additional hardware expenses and computational complexity for interference management. Alternatively, the co-design philosophy shares unitary hardware platform and transmit waveform for simultaneous sensing and communications, which reaches true harmony across space, time and spectrum domains. Therefore, this article is concentrated on the co-design DFRC.

\vspace{-2mm}
\subsection{Related Works}\label{S1.1}
 
An appropriate dual-functional waveform design is crucial\cite{Zhou_ojcoms_22}, which is challenging due to the diverse requirements of sensing and communication functions. Specifically, sequences with good auto-correlation property are usually preferable for sensing and positioning applications, while 
communication symbols tend to be random and stochastic, making it challenging to ensure consistent auto-correlation properties.
To address this problem, one promising representative is orthogonal frequency division multiplexing (OFDM) \cite{Zhou_ojcoms_22,Sturm_OFDM_2009,Braun_OFDM_2010}, which is widely adopted in current communication standards like the fourth-generation (4G) Long-Term Evolution LTE \cite{LTE_wc_10} and 5G new radio (NR) \cite{5GNR_twc_2021} thanks to the merits of OFDM signals, including robustness to frequency-selective fading and easy implementation. 
Despite of its randomness, the OFDM waveform possesses perfect auto-correlation property when using constant-amplitude constellations, e.g., phase-shift keying (PSK) \cite{Popovic_tit_18}. Under such a waveform design, OFDM can readily support accurate sensing/positioning with marginal modifications to the existing infrastructure \cite{2023arXiv230513924L}. However, such an ideal auto-correlation property is no longer guaranteed when the quadrature amplitude modulation (QAM) format is employed for higher spectral efficiency \cite{Popovic_tit_18,Zhou_taes_19}. Moreover, for the application scenario which performs sensing and positioning in a broadcasting/scanning mode, whilst providing directional access to user terminals simultaneously \cite{Dong_twc_23,Chen_icc_22}, the communication and sensing components should be non-overlapped in the time-frequency domain to avoid possible interference. 

To tackle the aforementioned issues, a dual-functional OFDM waveform design with interleaved subcarriers, abbreviated as OFDM-IS, was explored in the existing literature, where the dual functions are allocated with orthogonal spectrum resources  \cite{Bica_conf_19,Shi_sensorsJ_2019,Shi_systemsJ_21,Cao_wcnc_23,Chen_cl_23}. Specifically, in  \cite{Bica_conf_19}, the assignment and power allocation of OFDM subcarriers were optimized for the sensing and communication subsystems using a compound mutual-information (MI) based objective function, where both a radar-selfish and a balanced design strategies were developed. Another joint subcarrier and power allocation optimization strategy was proposed in \cite{Shi_sensorsJ_2019} and \cite{Shi_systemsJ_21}, aiming to minimize the total power consumption under constraints on the MI metric for radar sensing and the data rate for communications. The work \cite{Cao_wcnc_23} further proposed a robust multi-carrier waveform design against imperfect channel state information, where the bit and power allocation strategies were optimized with a greedy algorithm. 
Rather than adopting the MI-related metrics for radar sensing, the study \cite{Chen_cl_23} optimized the joint subcarrier and power allocation strategy between the dual functions by minimizing the sidelobe-to-peak ratio in the radar range profile\footnote{The radar range profile is referred to as a one-dimensional correlation function of the sensing sequence.}, whilst ensuring an acceptable level of the communication data rate. 

Unlike the aforementioned works which merely focused on the frequency- and power-domain characteristics, 
some studies extend the dual-functional waveform design to the time domain by involving multiple consecutive OFDM symbols \cite{m1,m2,m3,Ma_iotj_23}.
The work \cite{m1} optimized power allocation in a time-frequency range of interest to realize a favorable trade-off between sensing and communications with limited feedforward. In this study, however, the radar utilized the random and stochastic communication symbols for target detection, which limited the sensing performance.
Another power allocation optimization was proposed in \cite{m2}, aiming to maximize the compound signal-plus-distortion-and-noise ratio for sensing and communication subsystems. Since the time-frequency resources allocated for the two subsystems overlapped in the work \cite{m2}, there may be significant interference between them.
An optimum radar waveform was proposed in \cite{m3} to maximize the channel capacity by minimizing the distance between the communication symbols and the radar interference.
Although this design further enhanced the capacity of the communication system, it neglected the required properties for the radar sequence.
In \cite{Ma_iotj_23}, the optimum OFDM-IS design for dual-functional waveform is investigated in both time and frequency domains, and multiple resource element (RE) assignment strategies were proposed.
These strategies can achieve large time-frequency radar aperture with a tiny fraction of OFDM resources, whose optimality, however, was not validated with sufficient theoretical derivations. 

 \vspace{-2mm}
\subsection{Motivation and Our Contributions}\label{S1.2}
 \vspace{-1mm} 
From the above discussions, most of the existing literature on OFDM-based dual-functional waveform design only consider subcarrier assignment and power allocation within a single OFDM symbol. In order to improve the speed resolution in radar sensing, however, the coherent processing interval usually have to be extended by incorporating multiple consecutive OFDM symbols \cite{Sturm_procieee_11}. Considering the time-varying characteristics of the channel in high-speed scenarios, e.g., autonomous driving, it is difficult to apply the resource allocation strategy within one single OFDM symbol to the extended time interval. 
Consequently, to ensure the accuracy of positioning and speed estimation, optimization of the waveform design across multiple consecutive OFDM symbols is necessary. 
The existing attempts for multiple symbols design mostly fail to keep the communication and sensing signals orthogonal and non-interfering with each other, and consequently the diverse design requirements of the two subsystems cannot be fully satisfied.
To the authors' best knowledge, the work \cite{Ma_iotj_23} is the only existing reference involving multiple consecutive OFDM symbols that keeps the communication and sensing signals orthogonal by RE and subcarrier assignment, but the design of \cite{Ma_iotj_23} suffers from some drawbacks as discussed above. Against this background, we propose a cross-domain dual-functional waveform design based on the OFDM-IS structure. The waveform coefficients are optimized across time, frequency, power and even delay-Doppler domains, where both a communication-centric and sensing-centric design methodologies are developed, respectively. The main contributions of this paper can be summarized as follows.
 
\begin{enumerate}
\item \emph{Communication-centric waveform design}: 
To enhance the sensing performance whilst ensuring the optimal data rate,
a fraction of REs within each frame of multiple consecutive OFDM symbols are optimally assigned for communications through a water-filling algorithm based on prior knowledge of the time-frequency doubly-dispersive channel. The radar sensing components are then constituted by the concatenation of the remaining REs, where the energy budget is optimized in a subcarrier-wise manner to guarantee high peak-to-sidelobe ratio (PSLR) of the radar ambiguity function. The phase-frequency characteristic of the sensing component is further adjusted based on the branch-and-bound (BB) algorithm for the peak-to-average power ratio (PAPR) reduction.

\item \emph{Sensing-centric waveform design}: To ensure superior target sensing performance, we firstly construct a `locally' perfect auto-correlation property by shaping the radar ambiguity function of the integrated waveform within a pre-defined region of interest (RoI) in the delay-Doppler domain. Next we approximate Hermitian symmetry for the 2-dimensional ambiguity function. The unit cells of the ambiguity function beyond RoI, referred to as `irrelevant cells' for brevity, can then be directly manipulated to determine the power allocation pattern for sensing, where the REs with relatively low sensing power are employed for data transmission with water filling. Therein the irrelevant cells and the power allocation strategy for communications are jointly optimized for throughput enhancement in an alternating iterative manner.

\item Numerical results are provided to validate the superiority of the proposed communication-centric and sensing-centric waveform designs. The proposed communication-centric waveform is capable of achieving a high PSLR within RoI and a low PAPR while maintaining the optimal achievable data rate, compared with its classical counterparts. On the other hand, while ensuring a locally perfect auto-correlation property, the proposed sensing-centric waveform can approximate the maximum achievable data rate, demonstrating the feasibility of the proposed alternating iterative algorithm. Moreover, the effects of the key parameters on sensing and communication performance are investigated, to provide valuable guidance for the implementation of the proposed sensing-centric waveform design.
\end{enumerate}

\subsection{Structure and Notations}\label{S1.3}
The remainder of this paper is organized as follows. Section~\ref{S2} introduces the system model. Section~\ref{cc} defines the RoI in the ambiguity function and presents the proposed communication-centric waveform design. In Section~\ref{sc}, the sensing-centric waveform design is formulated as an optimization problem and an alternating iterative algorithm is developed to solve the problem. Numerical results are provided in Section~\ref{S5}, which is followed by conclusions in Section~\ref{S6}.

In this paper, DFT$(\mathbf{X},1)$ and DFT$(\mathbf{X},2)$ denote performing discrete Fourier transformation (DFT) on every row and every column of matrix $\mathbf{X}$, respectively. IDFT$(\mathbf{X},1)$ and IDFT$(\mathbf{X},2)$ denote performing inverse DFT (IDFT) on every row and every column of $\mathbf{X}$, respectively. $(\mathbf{X})^\mathrm{T}$ and $(\mathbf{X})^*$ stand for the transpose and conjugate of $\mathbf{X}$, respectively. $|\mathbf{X}|$ and $\text{angle}(\mathbf{X})$ denote the element-wise amplitude and phase values of $\mathbf{X}$, respectively. $|\mathbf{X}|^2$ is a matrix that contains the element-wise absolute square values of $\mathbf{X}$, and $\|\mathbf{X}\|$ denotes the Frobenius norm of $\mathbf{X}$. $\mathbf{X}\odot\mathbf{Y}$ denotes the Hadamard product of matrices $\mathbf{X}$ and $\mathbf{Y}$. For a two-dimensional matrix $\mathbf{X}$, $X(m,k)$ denotes the element in the $m$-th row and the $k$-th column of $\mathbf{X}$.  $\mathbf{1}$ denotes the all-one matrix. $x \sim \mathcal{CN}(0,\sigma^2)$ represents the random variable $x$ that follows a complex Gaussian distribution with mean $0$ and variance $\sigma^2$. $\lfloor\cdot \rfloor$ is the floor operation that rounds a real number to the nearest integer less than or equal to the number. $\text{card}({\cal N})$ denotes the cardinality of set ${\cal N}$.

\begin{figure}[!t]
\center
\includegraphics[width=0.95\linewidth, keepaspectratio]{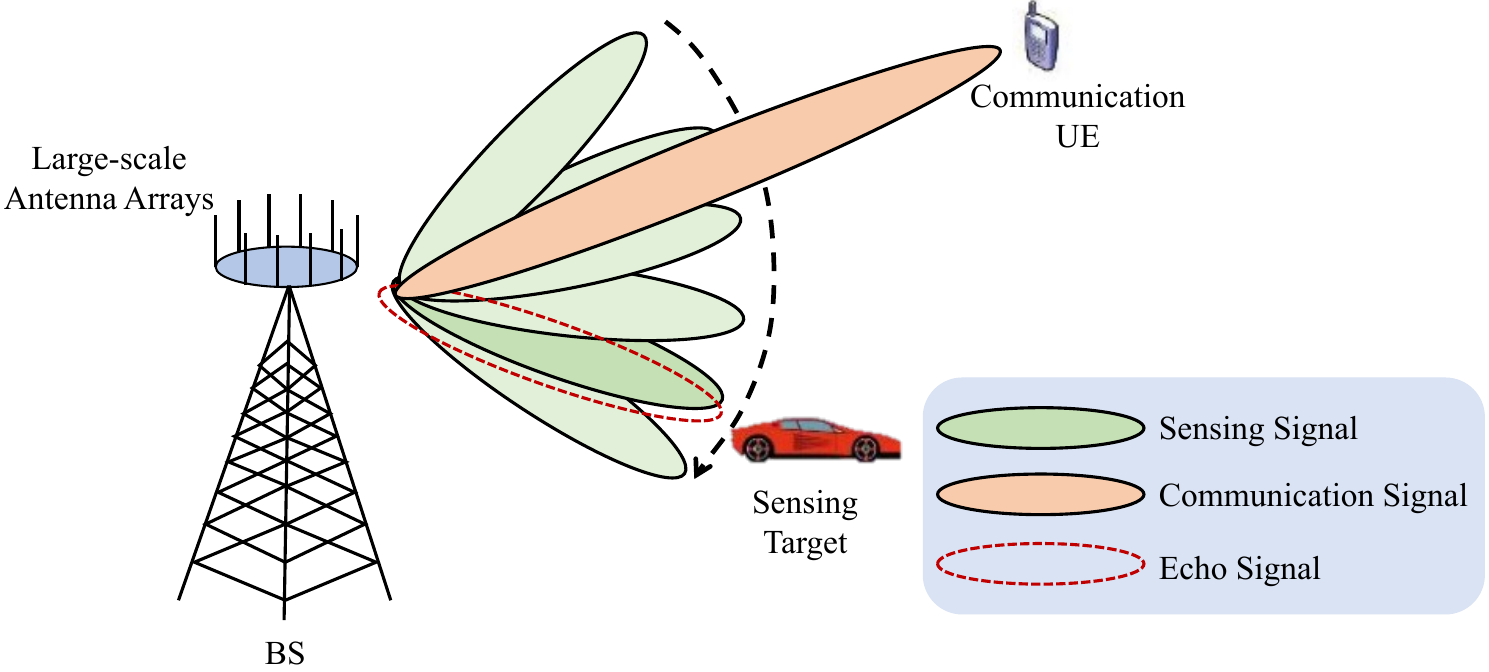}	
\vspace*{-2mm}
\caption{Typical wireless network-based monostatic radar sensing scenario with large-scale antenna arrays.}
\label{fig1}
\vspace*{-2mm}
\end{figure}

\section{System Model}\label{S2}
We consider a wireless network-based monostatic radar sensing scenario as illustrated in Fig.~\ref{fig1}, where a base station (BS) sends data to a user equipment (UE) while simultaneously scanning different directions alternatively for target sensing by emitting radar signals. To guarantee sufficient array gain, large-scale antenna arrays are employed to form highly directional beams for sensing and communications, respectively, especially for millimeter wave frequencies and above. Since the antenna array with directional beamforming is approximately equivalent to a single directional antenna, we consider single-input single-output systems for both sensing and communications for simplicity.

\begin{figure*}[!t]
\center
\includegraphics[width=0.75\linewidth, keepaspectratio]{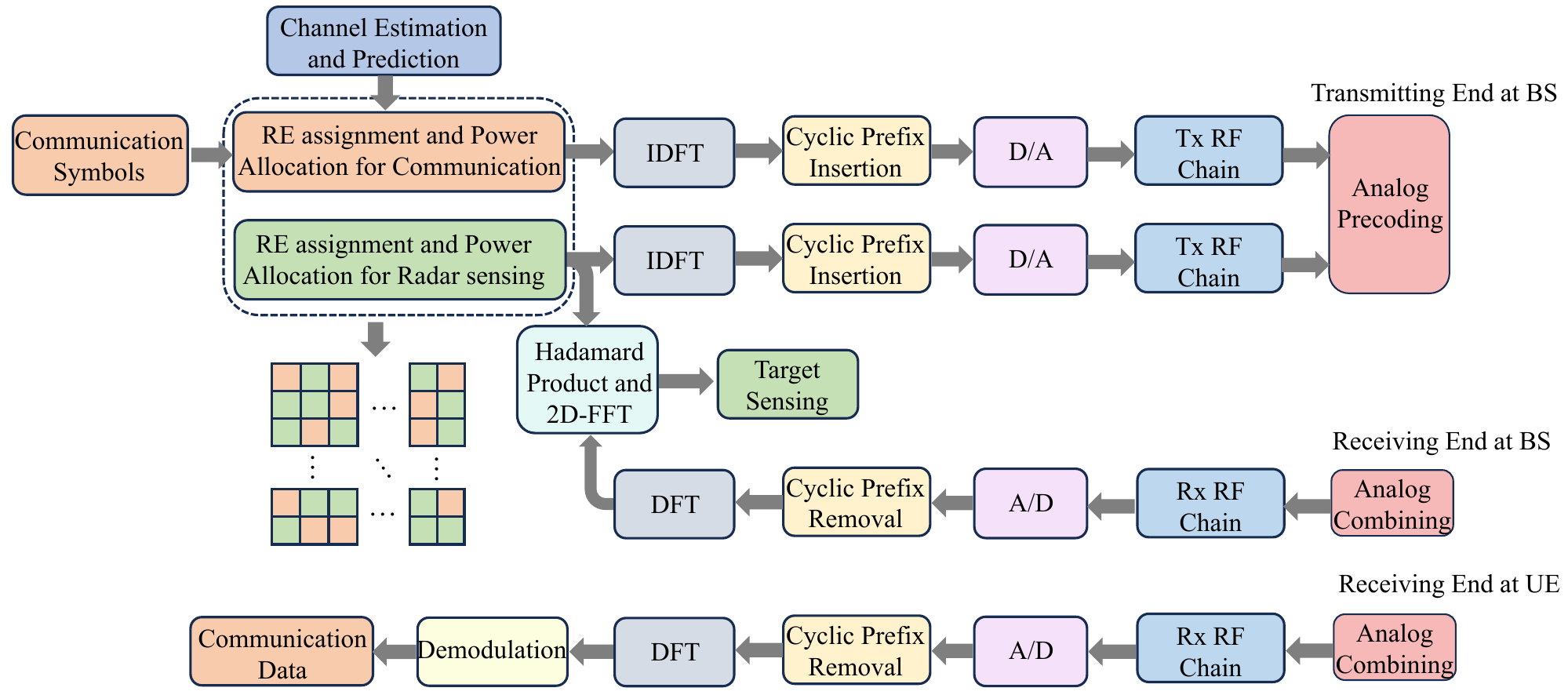}	
 \vspace*{-2mm}
\caption{Transceiver model of the OFDM-IS-based dual-functional system.}
\label{fig2}
\vspace*{-2mm}
\end{figure*}

\vspace{-4mm}
\subsection{Signal Model}\label{S2.1}
The OFDM-IS waveform is employed for target sensing and data transmission, and the corresponding transceiver model is depicted in Fig.~\ref{fig2}. Specifically, a frame of $M$ consecutive OFDM symbols with $N_{\rm{c}}$ subcarriers is considered for each coherent processing interval and the $k$-th subcarrier of the $m$-th symbol is called the $(m,k)$-th RE. To avoid mutual interference between the sensing and communication, these two subsystems occupy different REs in each frame \cite{Chen_cl_23,Shi_systemsJ_21,Bica_conf_19}. For clarity we employ the matrix $\mathbf{U} \in \mathbb{Z}^{M \times {N_c}}$ to indicate whether REs are selected for sensing or communication purpose
\begin{equation}\label{u1} 
	U(m,k) =
   \begin{cases}
   1, \, \text{$(m,k)$-th RE is for sensing,} \\
   0, \, \text{$(m,k)$-th RE is for communications.}
   \end{cases}\!\!
\end{equation}
Let $\mathbf{S}\! \in\! \mathbb{C}^{M \times {N_c}}$ be the transmit signal matrix with $S(m,k)$ representing the modulated symbol on the $(m,k)$-th RE. Then, the sensing and communication transmit signal matrices, denoted as $\mathbf{S}_{\rm{r}}\! \in\! \mathbb{C}^{M \times {N_c}}$ and $\mathbf{S}_{\rm{c}}\! \in\! \mathbb{C}^{M \times {N_c}}$, can be expressed as 
\begin{align} 
	\mathbf{S}_{\rm{r}} =& \mathbf{U} \odot \mathbf{S}, \label{eqSe} \\
	\mathbf{S}_{\rm{c}} =& (\mathbf{1}-\mathbf{U})\odot  \mathbf{S} . \label{eqCo}
\end{align}

Define $\mathbf{P}_{\rm{r}}\! \in\! \mathbb{R}^{M \times {N_c}}$ and $\mathbf{P}_{\rm{c}}\! \in\! \mathbb{R}^{M \times {N_c}}$ as the power allocation matrices for sensing and communications, which can be written as ${P}_{\rm{r}}(m,k)\! =\! |{S}_{\rm{r}}(m,k)|^2$ and ${P}_{\rm{c}}(m,k)\! =\! |{S}_{\rm{c}}(m,k)|^2$, respectively. By modulating $\mathbf{S}_{\rm{r}}$ and $\mathbf{S}_{\rm{c}}$ on different REs of the OFDM frame, the time-domain transmit signals, $x_{\rm{i}}(t)$, $\rm{i} \in \{\rm{r}, \rm{c}\}$, for sensing and communications, respectively, can be expressed as
\begin{align}\label{multi} 
	x_{\rm{i}}(t) \!=&\! \frac{1}{\sqrt{N_c}} \sum_{m=0}^{M-1} \! \sum_{k=0}^{N_c-1} S_{\rm{i}}(m,k) e^{\textsf{j}2\pi k \Delta f (t-mT)} \text{rect}\Big(\!\frac{t\!-\!mT}{T}\!\Big), 
\end{align}
where $\Delta f$ is the subcarrier spacing and $T=1/ \Delta f$ is the duration of one OFDM symbol, while $\text{rect}(\cdot)$ is the rectangle function, which is defined by
\begin{equation}\label{rect} 
	\text{rect}(t)=
   \begin{cases}
   1, \quad 0 \leq t < 1, \\
   0, \quad \text{otherwise}.
   \end{cases}
\end{equation}
Sampling $x_{\rm{i}}(t)$ with the period $T/N_c$ leads to the discrete sequence $\bar{x}_{\rm{i}}(n)=x_{\rm{i}}(nT/N_c)$ as:
\begin{align}\label{d_multi} 
	\bar{x}_{\rm{i}}(n) =& \frac{1}{\sqrt{N_c}} \sum_{m=0}^{M-1} \sum_{k=0}^{N_c-1} S_{\rm{i}}(m,k) e^{\textsf{j}2\pi \frac{k (n-mN_c)}{N_c}} \text{rect}_m(n),  
\end{align}
with $n \!=\! 0,1,\cdots,MN_c\!-\!1$, where $\text{rect}_m(n)\!= \text{rect}\big(\frac{n-mN_c}{N_c}\big)$. The cyclic prefix (CP) of length $T_{\rm{G}}$ is then inserted to mitigate the inter-frame interference. Therefore, the total duration of one OFDM symbol $T_{\rm{O}}$ is $T_{\rm{O}} = T + T_{\rm{G}}$. Afterwards, the discrete signals are fed into the digital-to-analog (D/A) converter and the transmitting (Tx) radio frequency (RF) chain. 
Finally, the directional beams obtained by the analog precoding are transmitted through the antennas for sensing and communications.

\vspace{-1mm}
\subsection{Communication Channel Model}\label{S2.2}
\vspace{-1mm}
The standard multi-path time-varying channel model is considered, which is expressed as
\begin{align}\label{eqh} 
	h_{\rm{c}}(t,f) = \sum_{l=1}^{L} \alpha_{l}e^{\textsf{j}2\pi(v_lt-\tau_lf)},
\end{align}
where $L$ is the number of paths, and $\alpha_{l}$, $\tau_l$ and $v_l$ denote the complex gain, delay and Doppler shift of the $l$-th path, respectively \cite{ref5}. For simplicity, we assume that each RE experiences time-invariant flat channel fading \cite{senol_tsp_21}. By sampling the channel with the period of $T_{\rm{O}}$ in time domain and $\Delta f$ in frequency domain, the discrete channel matrix $\mathbf{H}_{\rm{c}} \in \mathbb{C}^{M \times {N_c}}$ corresponding to different REs within the OFDM frame 
can be written as
\begin{align}\label{eqhh} 
	H_{\rm{c}}(m,k) = \sum_{l=1}^{L} \alpha_{l}e^{\textsf{j}2\pi(v_lmT_O-\tau_lk \Delta f)},
\end{align}
which is assumed to be perfectly estimated and predicted. At the receiving (Rx) end, after the analog combining, the Rx RF chain, the analog-to-digital (A/D) conversion, CP removal and DFT operation, the received signal matrix $\mathbf{Y}_{\rm{c}} \in \mathbb{C}^{M\times N_c}$ is obtained, which can be written as 
\begin{align}\label{eqRxS} 
  \mathbf{Y}_{\rm{c}} =& \mathbf{H}_{\rm{c}} \odot  \mathbf{S}_{\rm{c}} + \mathbf{W}_{\rm{c}}. 
\end{align}
Here $\mathbf{W}_{\rm{c}}\! \in\! \mathbb{C}^{M\times N_c}$ is the additive white Gaussian noise (AWGN) matrix whose entries follow $W_{\rm{c}}(m,k)\! \sim\! \mathcal{CN}(0,\sigma_{\rm{c}}^2)$. 
For demodulation, the UE first estimates the channel matrix based on the reference signals transmitted by the BS. According to the estimated channel matrix, the UE is able to deduce the RE assignment matrix $\mathbf{U}$ and acquire the indices of communication REs.
Then the communication data can be extracted from $\mathbf{Y}_{\rm{c}}$ by demodulation.

 \vspace{-1mm}
\subsection{Target Sensing}\label{S2.3}
 \vspace{-1mm}
In this paper, the target sensing process mainly includes the target detection, positioning and speed estimation. By denoting the distance and radial speed of the target relative to the BS as $d$ and $u$, respectively, the received echoes $\bar{y}_{\rm{r}}(n)$ of the sensing subsystem can be written as 
\begin{align}\label{senging} 
	\bar{y}_{\rm{r}}(n) =& \bar{x}_{\rm{r}}\bigg(n-\Big\lfloor\frac{2d}{cT_{\rm{s}}}\Big\rfloor\bigg)e^{\textsf{j}\big(2\pi n \frac{2uf_{\rm{c}}T_{\rm{s}}}{c}\big)}+ {w_{\rm{r}}}[n],
\end{align}
where $f_c$ and $T_{\rm s}$ denote the carrier frequency and the sampling period, respectively, while ${w_{\rm{r}}}[n]$ denotes the thermal noise plus the clutters from other directions, 
 satisfying $w_{\rm{r}}[n] \sim \mathcal{CN}(0,\sigma_{\rm{r}}^2)$. The sensing signal matrix $\mathbf{Y}_{\rm{r}} \in \mathbb{C}^{M \times {N_c}}$ can be derived by the DFT and serial-to-parallel operation on $\bar{y}_{\rm{r}}(n)$, and can be expressed as
\begin{equation}\label{eqSenM} 
 Y_{\rm{r}}(m,k) = \sum_{n=(m-1)N_{\rm{c}}}^{mN_{\rm{c}}-1}\bar{y}_{\rm{r}}(n)e^{\textsf{j}2\pi\frac{ (n- (m-1)N_{\rm{c}})k}{N_{\rm{c}}}}.
\end{equation}

Assuming that the time delay of the target is shorter than $T_{\rm{G}}$, the cross-correlation between the transmitted sensing sequence and its echoes is equivalent to performing DFT and IDFT on the Hadamard product of $(\mathbf{S}_{\rm{r}})^*$ and $\mathbf{Y}_{\rm{r}}$, which is expressed as
\begin{align}\label{eqSenCal} 
  \mathbf{E} =& \text{DFT}(\text{IDFT}((\mathbf{S}_{\rm{r}})^* \odot \mathbf{Y}_{\rm{r}},1),2).
\end{align}
Let $\nu = 0,1,\cdots,M-1$ and $\mu = 0,1,\cdots,N_{\rm{c}}-1$ denote the indices of $\mathbf{E}$ in the Doppler and delay domains, respectively.  The target can be detected by the hypothesis tests:
\begin{align}\label{eqHtest} 
	\text{H}_1:\frac{|E(\nu,\mu)|^2}{\theta(\nu,\mu)} >\Gamma,\quad \text{H}_0:\frac{|E(\nu,\mu)|^2}{\theta(\nu,\mu)} < \Gamma,
\end{align}
where $\Gamma$ is a predefined test threshold. Hypothesis $\text{H}_1$ represents that the point $(\nu,\mu)$ corresponds to a true target, and vice versa for hypothesis $\text{H}_0$. $\theta(\nu,\mu)$ is the average noise power estimation at the point $(\nu,\mu)$, which can be calculated by averaging the value of $|E(\nu,\mu)|^2$ \cite{ref6, zf}.

For a given point $(\nu_0,\mu_0)$, if hypothesis $\text{H}_1$ holds true, distance $d$ and radial speed $u$ of the corresponding target can be expressed respectively as \cite{Sturm_procieee_11}
\begin{align} 
  d =& \frac{c\mu_0}{2N_{\rm{c}}\Delta f}, \quad \mu_0 = 0,1,\cdots,N_{\rm{c}}-1,  \label{eqE-d} \\
  u =&
  \begin{cases}
   ~~\frac{c\nu_0}{2Mf_{\rm{c}}T_{\rm{O}}}, \quad  \nu_0 = 0,1,\cdots,\lfloor M/2 \rfloor, \\
   -\frac{c(M-\nu_0)}{2Mf_{\rm{c}}T_{\rm{O}}}, \quad \nu_0 = \lfloor M/2 \rfloor + 1,\cdots,M-1,
   \end{cases} \label{eqE-u}
\end{align}
where $c$ is the speed of light. Since the time delay of the target is limited by the CP length, the maximum sensing ranges of distance and speed can be expressed as $\big[0,\frac{cT_{\rm{G}}}{2}\big)$ and $\big(\!\!-\!\frac{c }{4f_{\rm{c}}T_{\rm{O}}},\frac{c }{4f_{\rm{c}}T_{\rm{O}}}\big)$, respectively. However, in practical applications, such as autonomous driving, the ranges of interest are usually smaller than the maximum sensing ranges. This will be discussed in the next section.

\vspace{-2mm}
\section{Communication-Centric Waveform Design}\label{cc} 

In this design, the sensing performance of the dual-functional waveform is improved without compromising the optimal communication data rate.
Firstly, a fraction of REs with good channel conditions are firstly assigned for communication to optimize the achievable data rate. Then the sensing energy budget is optimally allocated to the remaining REs to obtain high PSLR of the ambiguity function for sensing accuracy enhancement. Finally, the phases of sensing symbols are designed for PAPR reduction using the BB algorithm. Before detailing this design, we introduce the ambiguity function of the OFDM-based sensing waveform.

\vspace*{-3mm}
\subsection{Ambiguity Function of OFDM-Based Sensing Waveform}\label{S3.1}

The ambiguity function is defined as a two-dimensional correlation function in delay-Doppler domain, which can be categorized into the cross-ambiguity function and the auto-ambiguity function. We focus on the auto-ambiguity function of the sensing sequence, i.e., its auto-correlation property, which reflects its capability of radar sensing. 
Because the auto-ambiguity function is utilized to characterize the property of the sequence itself without considering external operations on the sequence, the cyclic prefix part is not included in the following derivation.
Specifically, the auto-ambiguity function $\chi_{a}(\nu,\mu)$ of the discrete signal $x(n)$ with length $N$ can be expressed as \cite{ref5, ref6}
\begin{align}\label{aaf} 
	\chi_{a}(\nu, \mu) =&  \sum_{n=0}^{N-1} x(n)x^{*}(n+\mu)e^{\textsf{j}2\pi \nu n/N},
\end{align}
where $\mu$ and $\nu$ denote the delay and Doppler indices, respectively. 
In the practical radar system, without consideration of noise, the cross-correlation matrix between the transmitted sensing sequence and its echoes is approximately equivalent to the ambiguity function shifted based on the time delay and Doppler frequency offset.
By substituting the OFDM-based sensing sequence (\ref{d_multi}) into (\ref{aaf}), the auto-ambiguity function can be derived as {\footnote{Note that due to the existence of CP, the correlation of intra-symbol ($m_1=m_2$) points is calculated as circular correlation.}}
\begin{align}\label{multi_auto} 
	& \chi_{a}(\nu,\mu) = \!\! \sum_{n=0}^{MN_c-1} \bigg(\!\sum_{m_1=0}^{M-1}\sum_{k_1=0}^{N_c-1}S(m_1,k_1) \psi(n,m_1,k_1) \!\bigg) \nonumber \\
  & \hspace*{10mm}\times\! \bigg(\!\!\sum_{m_2=0}^{M-1}\!\sum_{k_2=0}^{N_c-1}S^*(\!m_2,k_2\!) \psi^{*}(\!n\!+\!\mu,m_2,k_2\!) \!\!\bigg) e^{\textsf{j}2\pi \frac{\nu n}{ MN_c}} \nonumber \\
   & \hspace*{5mm} = \!\! \sum_{n=0}^{MN_c-1} \!\!\bigg( \!\sum_{m_1=0}^{M-1} \!\Big(\!\!\sum_{m_2=0}^{M-1} \psi(\!n,m_1,k_1\!)\psi^{*}(\!n\!+\!\mu,m_2,k_2\!) \!\Big) \nonumber \\
  & \hspace*{10mm} \times\! \Big(\!\sum_{k_1=0}^{N_c-1} \sum_{k_2=0}^{N_c-1} S(m_1,k_1)S^*(m_2,k_2) \!\Big)\! \bigg)e^{\textsf{j}2\pi  \frac{n \nu}{ MN_c}},  
\end{align}
where $\psi(n,m,k) = e^{\textsf{j}2\pi k (n-mN_c)/N_c}\text{rect}_{m}(n)$.
Due to the CP protection, the inter-frame ($m_1 \neq m_2$) interference is negligible. Therefore, (\ref{multi_auto}) can be further expressed as
\begin{align}\label{multi_auto1} 
	\chi_{a}(\nu,\mu) = & \sum_{m=0}^{M-1} \sum_{\bar{n}=0}^{N_c-1} \sum_{k_1=0}^{N_c-1} \sum_{k_2=0}^{N_c-1}
 \Big(e^{\textsf{j}2\pi \frac{(k_1-k_2)(m{N_c} + \bar{n})-k_2\mu}{N_c}} \times \nonumber\\
  & S(m,k_1)S^*(m,k_2)\Big) e^{\textsf{j}2\pi \nu \frac{m{N_c} + \bar{n}}{MN_c}} ,
\end{align}
where $\bar{n} = n - m{N_c}$. For simplicity, we focus on the auto-correlation component of each subcarrier ($k_1 = k_2$), while neglecting the cross-correlation component between different subcarriers ($k_1 \neq k_2$), since the cross-correlation is marginal due to the subcarrier orthogonality \cite{ref6}. Therefore, the auto-ambiguity function can be approximated as
\begin{align}\label{multi_auto2} 
	\chi_{a}(\nu,\mu) \approx &
  \gamma(\nu,\mu) \cdot \eta(\nu) ,
\end{align}
with
\begin{align} 
  \gamma(\nu,\mu) \triangleq & \sum_{m=0}^{M-1} \sum_{k=0}^{N_c-1} P(m,k)e^{-\textsf{j}2\pi \mu k/N_c}e^{\textsf{j}2\pi \nu m/M}, \label{gamma}
\end{align}
\begin{align} 
 \eta(\nu) \triangleq & \sum_{\bar{n}=0}^{N_c-1} e^{\textsf{j}2\pi \nu \bar{n}/(MN_c)}. \label{eta}
\end{align}

Note that only $\gamma(\nu,\mu)$ is related to the waveform design. Since $\gamma(\nu,\mu)$ reaches its maximum when $\nu$ and $\mu$ are multiples of $M$ and $N_{\rm{c}}$, respectively, we consider the region $(\nu,\mu) \in \Omega =\big[-\lfloor\frac{M}{2}\rfloor,\cdots,-1,0,1,\cdots,M-1-\lfloor\frac{M}{2}\rfloor\big] \times \big[-\lfloor\frac{N_{\rm{c}}}{2}\rfloor,\cdots,-1,0,1, \cdots,N_{\rm{c}}-1-\lfloor\frac{N_{\rm{c}}}{2}\rfloor\big]$ to avoid the ambiguity problem in target detection \cite{ref7}. As the sensing distance and the speed of interest in practical applications are commonly smaller than the maximum sensing ranges corresponding to region $\Omega$ in the ambiguity function, we define the RoI $\Omega_{\bm{s}}$ in the ambiguity function as 
\begin{align}\label{rof} 
  \Omega_{\bm{s}}\! =&\! \left[-\!\left\lfloor\!\frac{M}{2a}\!\right\rfloor\!,\cdots,\!-\!1,0,1,\cdots,\frac{M}{a} \!-\!1\!-\!\left\lfloor\!\frac{M}{2a}\!\right\rfloor\right]\! \nonumber\\
  & \times  \! \left[0,1,2, \cdots,\frac{N_{\rm{c}}}{b} -\!1\!-\! \left\lfloor\frac{N_{\rm{c}}}{2b}\right\rfloor\right] .
\end{align}
Given the required sensing scopes of distance $[0, d_0]$ and speed $[-u_0, u_0]$, $a$ and $b$ are chosen as the largest factors of $M$ and $N_{\rm{c}}$, respectively, such that $\big(-\frac{c }{4af_{\rm{c}}T_{\rm{O}}},\frac{c }{4af_{\rm{c}}T_{\rm{O}}}\big)$ and $\big[0, \frac{c }{2 b\Delta f}\big)$ contain $[-u_0, u_0]$ and $[0, d_0]$, respectively. That is, for $\Omega_{\bm{s}}$ of (\ref{rof}), the sensing ranges of distance and speed are $\big[0, \frac{c }{2 a \Delta f}\big)$ and $\big(-\frac{c }{4bf_{\rm{c}}T_{\rm{O}}},\frac{c }{4bf_{\rm{c}}T_{\rm{O}}}\big)$, respectively. Let $F(\Omega_{\bm{s}})$ be the highest sidelobe in RoI, namely,
\begin{equation}\label{sidelobe} 
  F(\Omega_{\bm{s}}) \!= \!\max \{|\chi_{a}(\!\nu,\mu\!)|:\!(\!\nu,\mu\!) \in \Omega_{\bm{s}}, ~  (\!\nu,\mu\!) \neq (0,0)\}.
\end{equation}
The PSLR within $\Omega_{\bm{s}}$ is $\frac{|\chi_{a}(0,0)|}{F(\Omega_{\bm{s}})}$, which is adopted to characterize the sensing performance.

\vspace*{-1mm}

\subsection{Waveform Design Methodology}\label{S3.2}

The communication-centric waveform design consists of the following three steps. 

\emph{\textbf{Step~1.~Communication RE and power allocation}:}
Given the total communication power $\bar{P}_{\rm{c}}$ and the estimated communication channel $\mathbf{\hat{H}}_{\rm{c}} \in \mathbb{C}^{M \times {N_c}}$, the communication power allocation strategy is to maximize the achievable data rate, which is formulated as 
\begin{equation}\label{eqOpP} 
\begin{array}{rcl}
  \mathcal{P}1: & \max\limits_{\mathbf{P}_{\rm{c}}} & \sum\limits_{m=0}^{M-1} \sum\limits_{k=0}^{N_{\rm{c}} -1}\log \Big(1 + \frac{P_{\rm{c}}(m,k)|\hat{H}_{\rm{c}}(m,k)|^2}{\sigma_{\rm{c}}^2}\Big) , \\ 
  & \text{s.t.} & \sum\limits_{m=0}^{M-1} \sum\limits_{k=0}^{N_{\rm{c}} -1}P_{\rm{c}}(m,k) = \bar{P}_{\rm{c}}, \\
  & & P_{\rm{c}}(m,k) \geq 0,
\end{array}
\end{equation}
with $m=0,1,\cdots,M-1$ and $k=0,1,\cdots,N_{\rm{c}}-1$.
 For brevity, the ranges of values for $m$ and $k$ are omitted in sequel. Problem $\mathcal{P}1$ can be solved by using the Karush-Kuhn-Tucker (KKT) conditions, and the optimal solution is given by \cite{ref8}
\begin{equation}\label{solution1} 
   P_{\rm{c}}(m,k) = \max \Big\{\frac{1}{\beta \ln2} - \frac{\sigma_{\rm{c}}^2}{|\hat{H}_{\rm{c}}(m,k)|^2}, 0\Big\},
\end{equation}
where $\frac{1}{\beta \ln2} = \frac{\bar{P}_{\rm{c}} + \sum_{(m,k)\in {\cal N}_{\rm e}}\frac{\sigma_{\rm{c}}^2}{|\hat{H}_{\rm{c}}(m,k)|^2} }{\text{card}({\cal N}_{\rm{e}})}$ and ${\cal N}_{\rm e}$ contains the indices of all the REs with non-zero power, i.e., $P_{\rm{c}}(m,k)>0$.

\emph{\textbf{Step~2.~Sensing RE and power allocation}:}
According to (\ref{solution1}), if the $(m,k)$-th RE is under poor channel condition, i.e., $|\hat{H}_{\rm{c}}(m,k)|^2$ is less than the threshold $\varsigma=\sigma_{\rm{c}}^2\beta \ln2 $, it will not be allocated for communication. By employing these inactivated REs for sensing purpose, the time-frequency resources for both subsystems will be completely orthogonal without inducing mutual interference in principle, while ensuring the optimal communication performance. Accordingly, the matrix $\mathbf{U}$ defined in (\ref{u1}) can be expressed as 
\begin{equation}\label{u2} 
  U(m,k)=
   \begin{cases}
   1, \quad \text{if } |\hat{H}_{\rm{c}}(m,k)|^2 \leq \varsigma, \\
   0, \quad \text{if } |\hat{H}_{\rm{c}}(m,k)|^2 > \varsigma.
   \end{cases}
\end{equation}
Note that when the communication channel is relatively flat-fading, the number of sensing REs may prove insufficient to support high-resolution sensing. In order to mitigate this issue, a minimum threshold is set for the number of sensing REs, which is denoted as $\hat{N}_r$. 
If the number of sensing REs (i.e., the number of non-zero elements in $\mathbf{U}$), denoted as ${N}_r$, is less than the threshold $\hat{N}_r$, $\hat{N}_r - {N}_r$ communication REs with the lowest channel gains are re-assigned for sensing.
Since the REs with good channel conditions are chosen for communication first, the sensing REs may be discontinuous in both time and frequency domains, causing possible high sidelobe in the ambiguity function.
To reduce the sidelobe level for enhancing sensing performance, the power allocation strategy for sensing REs is to maximize the PSLR within $\Omega_{\bm{s}}$, which is formulated as
\begin{equation}\label{eqSenPall} 
\begin{array}{rcl}
  \mathcal{P}2 \!:\!  & \max\limits_{\mathbf{P}_{\rm{r}}} & \frac{|\chi_{a}(0,0)|}{F(\Omega_{\bm{s}})} , \\ 
  & \text{s.t.} &  \!\sum\limits_{m=0}^{M-1} \sum\limits_{k=0}^{N_c -1}P_{\rm{r}}(m,k) \! =  \!\bar{P}_{\rm{r}}, ~ P_{\rm{r}}(m,k) \! \geq \!0 ,
\end{array}
\end{equation}
where $\bar{P}_{\rm{r}}$ is the total sensing power budget. Since the main peak $|\chi_{a}(0,0)|$ is always equal to the total sensing power $N_{\rm{c}}\bar{P}_{\rm{r}}$, the problem $\mathcal{P}2$ is equivalent to minimize the highest sidelobe level given by:
\begin{equation}\label{eqSenPall2} 
\begin{array}{rcl}
  \mathcal{P}3\!: \!& \min\limits_{\mathbf{P}_{\rm{r}}} & \max\limits_{(0,0) \neq (\nu,\mu) \in \Omega_{\bm{s}}} |\chi_{a}(\nu,\mu)|,  \\ 
  & \text{s.t.} & \!\sum\limits_{m=0}^{M-1} \sum\limits_{k=0}^{N_c -1}P_{\rm{r}}(m,k)\! =\! \bar{P}_{\rm{r}}, ~ P_{\rm{r}}(m,k)\! \geq\! 0.
\end{array}
\end{equation}
 
Due to the minimax form of the objective function, it is difficult to solve problem $\mathcal{P}3$ directly.
To address this issue, a slack variable $z_0$ is introduced as the optimization objective. Accordingly, we add an extra constraint that all the sidelobes within $\Omega_{\bm{s}}$ are less than $z_0$. Then the optimization problem can be expressed as
\begin{equation}\label{p4} 
\begin{array}{rcl}
\mathcal{P}4:
  & \min\limits_{\mathbf{P}_{\rm{r}}} & z_0 , \\ 
  & \text{s.t.} & \!|\chi_{a}(\nu,\mu)|<z_0, ~ (0,0) \neq (\nu,\mu)\in \Omega_{\bm{s}}, \\
  & & \!\sum\limits_{m=0}^{M-1} \sum\limits_{k=0}^{N_c -1}\!P_{\rm{r}}(m,k)\! = \! \bar{P}_{\rm{r}}, ~ P_{\rm{r}}(m,k) \!\geq \!0.
\end{array}
\end{equation}
Problem $\mathcal{P}4$ now can be formulated into a quadratic problem, which is solvable using the CVX toolbox \cite{ref11}. 

\emph{\textbf{Step~3.~PAPR reduction}:}
In addition to the PSLR in the ambiguity function, the PAPR is also a significant metric for sensing sequences. A high PAPR results in nonlinear distortion of high power amplifier, which is detrimental especially for the millimeter wave and Terahertz bands \cite{ref12}. To tackle this problem, we need to minimize the PAPR of the sensing sequence, whilst retaining the high PSLR property. Recall that the PSLR is mainly related to the power allocation $\mathbf{P}_{\rm{r}}$ according to (\ref{multi_auto2}), which inspires us to minimize the PAPR by optimizing the phase of each element of $\mathbf{S}_{\rm{r}}$. Let $\theta(m,k)$ denote the phase of the symbol on the $(m,k)$-th RE. 
The PAPR of the sensing sequence among $M$ OFDM symbols can be expressed as 
\begin{align}\label{eqPAPR} 
  \text{PAPR} \!=\!& \frac{\max\limits_{m,n} \!\Big| \! \sum\limits_{k=0}^{N_{\rm{c}}-1}\!U(\!m,k\!)\sqrt{{P}_{\rm{r}}(\!m,k\!)}e^{\textsf{j}(\!\theta(\!m,k\!) \!+\!2\pi n k/N_{\rm{c}}\!)} \!\Big|^2}{\frac{1}{M}\sum\limits_{m=0}^{M-1}\sum\limits_{k=0}^{N_{\rm{c}}-1}U(m,k){P}_{\rm{r}}(m,k)}, 
\end{align} 
where $0 \leq m \leq M-1$ and $0 \leq n \leq N_c-1$. Since the average power of all the symbols has been determined by the proposed RE assignment and power allocation strategy, minimizing the PAPR is equivalent to minimizing the peak. Considering that the peak among $M$ symbols is the maximum value among the peak values of all symbols, we can reduce it by lowering the peak value of each symbol individually.
Additionally, the peak values corresponding to different OFDM symbols are independent of each other, the phase-frequency characteristics of different OFDM symbols can be optimized separately. Consider the case that the phases of all the modulated symbols on different sensing REs are selected from the set $\{0, 2 \pi/R, \cdots, 2 \pi(R-1)/R\}$ with the size $R$. Then the PAPR reduction problem for the $m$-th OFDM symbol can be formulated as 
\begin{equation}\label{eqPAPRop} 
\begin{array}{rcl}
  \mathcal{P}5: \!\!\!\!& \min\limits_{\pmb{\theta}_m} & \!\!\!\max\limits_{  n } \!\Big| \!\sum\limits_{k=0}^{N_{\rm{c}}-1}U(\!m,k\!)\sqrt{{P}_{\rm{r}}(\!m,k\!)}e^{\textsf{j}\theta(\!m,k\!)} e^{-\textsf{j}2\pi n k/N_{\rm{c}}} \!\Big|^2 , \\ 
  \!\!\!\!& \text{s.t.} & \!\!\!\theta(m,k) \in \{0, 2 \pi/R, \cdots, 2 \pi(R-1)/R\} ,
\end{array}
\end{equation}
where $\pmb{\theta}_m=\big[\theta(m,0),\theta(m,1),\cdots,\theta(m,N_{\rm{c}}-1)\big]^\mathrm{T}$. 

To solve this discrete optimization problem, we propose a heuristic searching method based on the BB algorithm to obtain a near optimal solution. The BB algorithm divides the whole feasible region into several sub-regions, which correspond to several sub-problems. For each sub-problem, well-designed functions are employed to estimate its upper and lower performance bounds. As the number of iterations increases, the minimum upper and lower bounds among all the sub-problems are obtained and updated. When the difference between the minimum upper and lower bounds is lower than a threshold, the iteration terminates and the solution that achieves the minimum upper bound is adopted as the final solution \cite{ref13}.

Let $\mathbf{A}_m$ denote $e^{\textsf{j}\pmb{\theta}_m}$, i.e., the $k$-th element of $\mathbf{A}_m$ is ${A}_m(k) = e^{\textsf{j} \theta(m,k)}$. Since the feasible region of ${A}_m(k)$ is $\Phi = \big\{1, e^{\textsf{j}\frac{2\pi}{R}}, \cdots, e^{\textsf{j}\frac{2\pi(R-1)}{R}}\big\}$, the whole feasible region of $\mathbf{A}_m$ is $\Psi^{(0)} = \Phi ^{N_{\rm{c}}}$, which is the Cartesian product of $N_{\rm{c}}$ $\Phi$s. The problem $\mathcal{P}5$ can be compactly written as
\begin{equation}\label{eqphiO} 
\begin{array}{rcl}
  \mathcal{P}(\Psi^{(0)}): & \min\limits_{\mathbf{A}_m} & f(\mathbf{\mathbf{A}_m}) , \\ 
  & \text{s.t.} & \mathbf{\mathbf{A}_m} \in  \Psi^{(0)},
\end{array}
\end{equation}
with 
\begin{align} 
f(\!\mathbf{A}_m\!) \!=\! \max_{ n  }\Big|\! \sum_{k=0}^{N_{\rm{c}}\!-\!1}U(\!m,k\!)\sqrt{\!{P}_{\rm{r}}(\!m,k\!)} {A}_m(\!k\!) e^{\!-\!\textsf{j}2\pi \frac{n k}{N_{\rm{c}}}}\!\Big|^2.
\end{align}

For each sub-region $\Psi\in \Psi^{(0)}$, denote its corresponding sub-problem as $\mathcal{P}(\Psi)$. A lower bound of $\mathcal{P}(\Psi)$ can be derived by a bounding function, which can be expressed as 
\begin{equation}\label{lb} 
  f_{\rm{L}}(\Psi) = f\big(\mathbf{A}_m^L\big),
\end{equation}
where $\mathbf{A}_m^L$ is a relaxed solution of $\mathcal{P}(\Psi)$ that achieves the lower bound. More  specifically, suppose that the first $k_0$ elements in $\Psi$ remain in their original values, and its $(k_0+1)$-th element is fixed to a feasible value. By extending the feasible region of the other elements to a continuous search region, a relaxed problem $\mathcal{P}_{\rm{L}}(\Psi)$ can be expressed as
\begin{equation}\label{eq.LB} 
\begin{array}{rcl}
 \mathcal{P}_{\rm{L}}(\Psi): & \min\limits_{{A}_m(k_0+1), \cdots,{A}_m(N_{\rm{c}}-1)} & f(\mathbf{\mathbf{A}_m}), \\ 
 & \text{s.t.} &  \big|A_m(k)\big|^2 \leq 1, 
\end{array}
\end{equation}
with $k = k_0+1,k_0+2,\cdots,N_{\rm{c}}-1$.
$\mathbf{A}_m^L$ can be derived by solving $\mathcal{P}_{\rm{L}}(\Psi)$ via the CVX toolbox, yielding the lower bound $f_{\rm{L}}(\Psi)$ of each sub-problem. On the other hand, by projecting $\mathbf{A}_m^L$ onto its closest feasible point, i.e.,
\begin{equation}\label{eq.UB} 
  \mathbf{A}_m^U = \argmin_{\mathbf{A}_m \in \Psi^{(0)}}||\mathbf{A}_m- \mathbf{A}_m^L||,
\end{equation}
an upper bound of each sub-problem is obtained, which can be written as 
\begin{equation}\label{ub} 
  f_{\rm{U}}(\Psi) = f(\mathbf{A}_m^U). 
\end{equation}

\begin{algorithm}[!t]
\caption{PAPR reduction method based on BB algorithm}
\label{alg1}
\begin{algorithmic}[1]
	\REQUIRE Sensing power allocation $\mathbf{P}_{\rm{r}}$, indicating matrix $\mathbf{U}$, termination threshold $\epsilon$, pruning threshold $N_{\rm{s}}$;
  \STATE Initialization: $\mathcal{S}=\{\mathcal{P}(\Psi^{(0)})\}$, $B_{\rm{L}}=f_{\rm{L}}(\Psi^{(0)})$, $B_{\rm{U}}=f_{\rm{U}}(\Psi^{(0)})$;
  \WHILE{$B_{\rm{U}}-B_{\rm{L}}>\epsilon$ and $\mathcal{S} \neq \varnothing$}
	  \STATE \textbf{Branching:} Select $\mathcal{P}(\Psi) \in \mathcal{S}$ such that $f_{\rm{L}}(\Psi)$ is the smallest; 
    \IF{$\Psi$ is non-divisible}
      \STATE Delete $\mathcal{P}(\Psi)$ from $\mathcal{S}$; 
			\STATE Continue; (go to line 3 \textbf{Branching})
    \ENDIF
    \STATE Partition $\Psi$ into $R$ sub-regions $\Psi_1,\Psi_2,\cdots,\Psi_{R}$ according to (\ref{partition});
    \STATE Delete $\mathcal{P}(\Psi)$ from $\mathcal{S}$;
    \STATE Add $\mathcal{P}(\Psi_1)$, $\mathcal{P}(\Psi_2),\cdots,\mathcal{P}(\Psi_{R})$ to $\mathcal{S}$;
		\STATE \textbf{Bounding:} According to (\ref{lb}) and (\ref{ub}), calculate the lower and upper bounds for $\mathcal{P}(\Psi_r)$, $r=1,2,\cdots,R$;
    \STATE Update $B_{\rm{U}}$ and $B_{\rm{L}}$;
    \STATE \textbf{Pruning:} If the lower bound of $\mathcal{P}(\Psi_r)$ is larger than $B_{\rm{U}}$, delete $\mathcal{P}(\Psi_r)$ from $\mathcal{S}$;
    \WHILE{ $\text{card}(\mathcal{S}) > N_{\rm{s}}$}
      \STATE Delete the sub-problem with the largest lower bound from $\mathcal{S}$;
    \ENDWHILE
  \ENDWHILE
  \RETURN $\pmb{\theta}_m \!=\! \text{angle}(\mathbf{A}_m^{\rm opt})$ with   $f(\mathbf{A}_m^{\rm opt})\! = \!B_{\rm{U}}$.
\end{algorithmic}
\end{algorithm} 

The proposed PAPR reduction procedure is summarized in Algorithm~\ref{alg1}. We initialize the problem set $\mathcal{S}$ as $\{\mathcal{P}\big(\Psi^{(0)}\big)\}$. The minimum upper bound $B_{\rm{U}}$ and lower bound $B_{\rm{L}}$ are initialized as the upper and lower bounds of $\mathcal{P}\big(\Psi^{(0)}\big)$, respectively. At each iteration, the sub-problem ${\cal P}(\Psi)$ with the smallest lower bound in $\mathcal{S}$ is selected. Note that the first $k_0$ elements of $\mathbf{A}_m$ in $\Psi$ have been set to fixed values in the previous iterations. If $k_0=N_{\rm c}$, i.e., all the elements of $\mathbf{A}_m$ have already been set to fixed values, $\Psi$ is non-divisible and ${\cal P}(\Psi)$ is deleted from $\mathcal{S}$, and the algorithm considers the next smallest lower-bound subproblem in $\mathcal{S}$. Otherwise, by denoting the fixed value of ${A}_m({k})$ as ${a}_{{k}} \in {\Phi}$, the region $\Psi$ can be expressed as $\Psi = \{{a}_{0}\} \otimes \{{a}_{1}\} \otimes \cdots \otimes \{{a}_{k_0-1}\}\otimes \Phi^{N_{\rm{c}}\!-k_0} \triangleq \Psi^{[k_0]} \otimes \Phi^{N_{\rm{c}}\!-k_0}$, where $\otimes$ denotes the Cartesian product. $\Psi$ is divided into $R$ smaller regions by fixing the $(k_0+1)$-th element of $\mathbf{A}_m$ to each of the $R$ values in $\phi$, while the feasible region of the remaining elements of $\mathbf{A}_m$ is the full search region $\Phi$, which can be expressed as 
\begin{align}\label{partition} 
  \begin{array}{l}
	\Psi_1 = \Psi^{[k_0]} \otimes \{1\} \otimes \Phi^{N_{\rm{c}}\!-\!k_0\!-\!1}, \\
  \Psi_2 =\Psi^{[k_0]} \otimes \{e^{\textsf{j}\frac{2\pi}{R}}\} \otimes \Phi^{N_{\rm{c}}\!-\!k_0\!-\!1}, \\
  \cdots, \\
  \Psi_{R} = \Psi^{[k_0]} \otimes \{e^{\textsf{j}\frac{2\pi(R\!-\!1)}{R}}\} \otimes \Phi^{N_{\rm{c}}\!-\!k_0\!-\!1}.
	\end{array}
\end{align}
Their corresponding problems, $\mathcal{P}\big(\Psi_1\big)$, $\mathcal{P}\big(\Psi_2\big),\cdots,\mathcal{P}\big(\Psi_{R}\big)$, are added to $\mathcal{S}$ and the problem $\mathcal{P}\big(\Psi\big)$ is deleted from $\mathcal{S}$. The lower and upper bounds of each of these created sub-problems are calculated by bounding functions to update $B_{\rm{L}}$ and $B_{\rm{U}}$. If the lower bound of a certain sub-problem is larger than $B_{\rm{U}}$, it is deleted from $\mathcal{S}$. To reduce the worst-case computational complexity, we also keep $\text{card}(\mathcal{S})$ to no more than a threshold $N_{\rm{s}}$ by deleting sub-problems with relatively larger lower bound from $\mathcal{S}$. When the difference between $B_{\rm{L}}$ and $B_{\rm{U}}$ is less than a threshold $\epsilon$ or $\mathcal{S}$ is an empty set, the iteration procedure terminates. The solution corresponding to $B_{\rm{U}}$ is the final solution $\mathbf{A}_m^{\rm opt}$, and the symbol phase $\pmb{\theta}_m$ is set as $\text{angle}(\mathbf{A}_m^{\rm opt})$.
As it can be seen, the worst-case computational complexity of the proposed PAPR reduction is $\mathcal{O}(\text{card}(\mathcal{S}){N_{\rm{c}}})$, which is much lower than the computation complexity of exhaustive searching ($\mathcal{O}(R^{N_{\rm{c}}})$). 
Besides, the computational complexity can be further reduced by adjusting the parameter $\epsilon$.

\vspace{-1mm}
\section{Sensing-Centric Waveform Design}\label{sc} 
\vspace{-1mm}
In this design, we first adjust the unit cells of the ambiguity function within its RoI to guarantee the `locally' perfect auto-correlation property. Based on the correspondence between the main part of the ambiguity function and the sensing power allocation strategy, the irrelevant cells beyond RoI determine the power allocation strategy. Then the irrelevant cells are optimized together with the communication power allocation strategy for throughput enhancement.

\vspace{-1mm}
\subsection{Problem Formulation}\label{S4.1}
\vspace{-1mm}
To provide accurate sensing, the sidelobe level in the ambiguity function can be adopted to characterize sensing performance, as mentioned in Section~\ref{cc}. A sensing sequence has a locally perfect auto-correlation property, when the value of the highest sidelobe within RoI in its ambiguity function is zero, which can be expressed as
\begin{align}\label{eqLZ} 
  |\chi_{a}(\nu,\mu)| =& \gamma(\nu,\mu) \eta(\nu) =0, \quad (0,0) \neq (\nu,\mu)\in \Omega_{\rm{s}}.
\end{align}
Since $\eta(\nu) \neq 0$ (when $m>1$), $|\chi_{a}(\tau,\nu)|=0$ is equivalent to $\gamma( \nu,\mu)=0$.
Given the total sensing power $\sum_{m=0}^{M-1}\sum_{k=0}^{N_{\rm{c}}-1}P_{\rm{r}}(m,k) = \bar{P}_{\rm{r}}$, $\gamma(0,0)$ is always equal to $\bar{P}_{\rm{r}}$. Therefore, the locally perfect auto-correlation property can be obtained by setting the function $\gamma( \nu,\mu)$ within RoI as
\begin{equation}\label{gamma1} 
	\gamma( \nu,\mu) =
    \begin{cases}
    \bar{P}_{\rm{r}}, \quad (\nu,\mu) = (0,0), \\
    0, \quad  (0,0)  \neq (\nu,\mu)\in \Omega_{\rm{s}}.
    \end{cases}
\end{equation}

According to (\ref{gamma}), $\gamma( \nu,\mu)$ is derived by performing DFT and IDFT on ${P}_{\rm{r}} (m,k)$ along its column and row, respectively. Therefore, ${P}_{\rm{r}}(m,k)$ can be derived by performing inverse operations on $\gamma( \nu,\mu)$, which can be expressed as 
\begin{align}\label{inverse} 
  {P}_{\rm{r}}(\!m,k\!) \!=\! & \frac{1}{MN_{\rm{c}}} \!
  \sum_{\nu= \!-\!\lfloor\! M/2\! \rfloor}^{M\!-\!1 \!-\!\lfloor\! M/2\!\rfloor} \! \sum_{\mu=\!-\!\lfloor\! N_{\rm{c}}/2\! \rfloor}^{N_{\rm{c}}\!-\!1 \!-\!\lfloor\! N_{\rm{c}}/2\!\rfloor}
   \gamma(\!\nu,\mu\!) e^{\textsf{j}2\pi \frac{\mu k}{N_{\rm{c}}} }e^{-\textsf{j}2\pi \frac{ \nu m}{M}}
 .
\end{align}
Since there is a one-to-one mapping between $\gamma(\nu,\mu)$ and ${P}_{\rm{r}}(m,k)$, we can use $\gamma( \nu,\mu)$ as the optimization variable instead of ${P}_{\rm{r}}(m,k)$ to reduce the dimension of the optimization variable. This is because for a sensing with a locally perfect auto-correlation property, the unit cells of $\gamma(\nu,\mu)$ within $\Omega_{\rm{s}}$ are fixed and we only need to optimize the unit cells in  the complementary set of $\Omega_{\rm{s}}$, which are called irrelevant cells.  ${P}_{\rm{r}}(m,k)$ is always a non-negative real number, the irrelevant cells should be designed under this constraint, which is formulated as
\begin{equation}\label{p8} 
\begin{array}{rcl}
  \mathcal{P}8: & \text{find} & {\gamma( \nu,\mu), \text{ for } (\nu,\mu) \in \bar{\Omega}_{\rm{s}}} , \\ 
  & \text{s.t.} & {P}_{\rm{r}}(m,k) \geq 0, \\
  & & {P}_{\rm{r}}(m,k) = P^*_r(m,k),
\end{array}
\end{equation}
where $\bar{\Omega}_{\rm{s}}$ denotes the complementary set of ${\Omega}_{\rm{s}}$ in $\Omega$, and ${P}_{\rm{r}}(m,k) = P^*_r(m,k)$ is equivalent to $\gamma( \nu,\mu)$ being centrohermitian symmetric, which can be expressed as 
\begin{equation}\label{symmetry} 
  \gamma( \nu,\mu) = \gamma^*( -\nu, -\mu), ~ (\nu,\mu) \in \Omega.
\end{equation}

\begin{figure*}[!t]
\vspace*{-1mm}
\center
\includegraphics[width=0.65\linewidth, keepaspectratio]{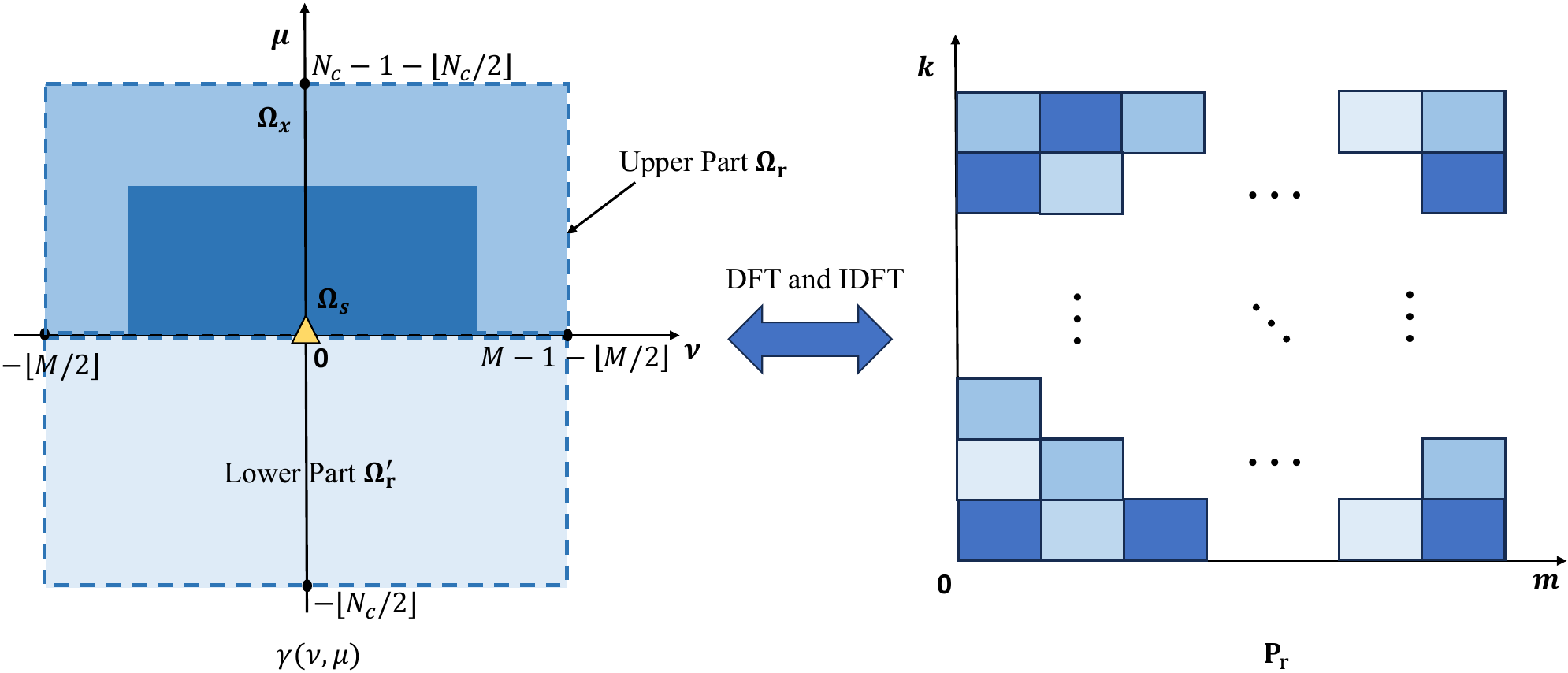}	
\vspace*{-2mm}
\caption{One to one mapping between $\gamma( \nu,\mu)$ and $\mathbf{P}_{\rm{r}}$. The left figure illustrates different regions of $\gamma( \nu,\mu)$, and the right figure shows the sensing power allocation $\mathbf{P}_{\rm{r}}$, where the depth of the color indicates the amount of power for the $(m,k)$-th RE.}
\label{fig_region} 
\vspace{-4mm}
\end{figure*}

As seen in Fig.~\ref{fig_region}, the upper part ${\Omega}_{\rm{r}}$ and lower part ${\Omega}^{'}_{\rm{r}}$ of the ambiguity function divided by the $\nu$-axis are mutually centro symmetric. Therefore, we only need to adjust the upper part denoted by ${\Omega}_{\rm{r}} = [-\lfloor M/2 \rfloor,-\lfloor M/2 \rfloor+1,\cdots, M -1-\lfloor M/2 \rfloor] \times [0,1,\cdots,N_{\rm{c}} -1-\lfloor N_{\rm{c}}/2\rfloor ]$. ${\Omega}_{\rm{r}}$ consists of the RoI ${\Omega}_{\rm{s}}$ and the outer region ${\Omega}_{\rm{x}}$. Since the unit cells in ${\Omega}_{\rm{s}}$ are fixed to ensure perfect auto-correlation property locally, the unit cells in ${\Omega}_{\rm{x}}$ are finally employed as the optimization variables, which can be expressed as
\begin{equation}\label{p9} 
\begin{array}{rcl}
  \mathcal{P}9: & \text{find} & {\gamma( \nu,\mu), \text{ for } (\nu,\mu) \in  {\Omega}_{\rm{x}}}, \\ 
  & \text{s.t.} & P_{\rm{r}}(m,k) \geq 0 .
\end{array}
\end{equation}
Specifically, ${\Omega}_{\rm{x}}$ can be expressed as ${\Omega}_{\rm{r}}/{\Omega}_{\rm{s}}$, where $/$ denotes the difference operation between sets. 
For brevity, $(\nu,\mu) \in  {\Omega}_{\rm{x}}$ is omitted below, i.e., the optimization of $\gamma( \nu,\mu)$ is referred to as the optimization of the unit cells of $\gamma( \nu,\mu)$ in ${\Omega}_{\rm{x}}$.
Each solution to $\mathcal{P}9$ corresponds to one of the possible realizations of $\mathbf{P}_{\rm{r}}$, which can all realize locally perfect auto-correlation property to guarantee superior sensing performance. The REs with relatively low sensing power budget, i.e., lower than a predefined threshold $\delta$, are considered to have a marginal impact on the sensing task. Hence, these REs are assigned for data transmission, where the indicating matrix $\mathbf{U}$ can be written as
\begin{equation}\label{u3} 
  U(m,k)=
    \begin{cases}
    1, \quad \text{if } {P}_{\rm{r}}(m,k) > \delta, \\
    0, \quad \text{if } {P}_{\rm{r}}(m,k) \leq \delta.
    \end{cases}
\end{equation}

Afterward, the power allocation for different communication REs is optimized for achievable data rate maximization. In this way, the achievable data rate is related to both the irrelevant cells of $\gamma( \nu,\mu)$ and the communication power strategy. The jointly optimization can be derived by combining $\mathcal{P}9$ and communication power allocation, which can be expressed as
\begin{equation}\label{p11} 
\begin{array}{rcl}
  \mathcal{P}11: \!\!\!\!
  & \!\!\!\! \max\limits_{\gamma(\!\nu,\mu\!),\mathbf{P}_c} & \!\! \!\!\!\sum\limits_{m=0}^{M-1}\sum\limits_{k=0}^{N_c -1}\log\Big(\!1 \!+\! \frac{(1\!-\!U(m,k)){P}_{\rm{c}}(m,k)|\hat{H}_{\rm{c}}(\!m,k\!)|^2}{\sigma_{\rm{c}}^2}\!\Big) , \\ 
  & \text{s.t.} & \!\!\!\!\sum\limits_{m=0}^{M-1} \sum\limits_{k=0}^{N_c -1}{P}_{\rm{c}}(m,k) = \bar{P}_{\rm{c}}, ~ P_{\rm{c}}(m,k) \geq 0, \\
  & & \!\!\!\!P_{\rm{r}}(m,k) \geq 0, \\
\end{array}
\end{equation}
where the relationship between $\gamma(\nu,\mu)$ and $U(m,k)$ is shown in (\ref{inverse}) and (\ref{u3}). Also the value for $\gamma(\nu,\mu)$ within RoI should satisfy (\ref{gamma1}), which is another constraint of the problem $\mathcal{P}11$.
One intuitive method of solving problem $\mathcal{P}11$ is to exhaustively search for the values of $\gamma( \nu,\mu)$ within ${\Omega}_{\rm{x}}$, which however imposes considerable computational complexity. On the other hand, since there are two types of optimization variables and the last constraint of the problem is non-convex, it is difficult to solve the problem via the KKT conditions directly. We propose a low-complexity alternating optimization algorithm to solve problem $\mathcal{P}11$ with near optimal solution.

\vspace*{-1mm}
\subsection{Alternating Optimization Algorithm}\label{S4,2}

$\mathcal{P}11$ can be naturally divided into two sub-problems, the communication power allocation problem and the irrelevant cell design problem, which corresponds to the optimization of $\mathbf{P}_{\rm{c}}$ and the irrelevant cells of $\gamma( \nu,\mu)$, respectively. For tractability of solving $\mathcal{P}11$, we optimize these two variables alternatively in an iterative manner. Denote the variables optimized after the $i$-th iteration as $\mathbf{P}^{(i)}_{\rm{c}}$ and $\gamma^{(i)}(\nu,\mu)$. Accordingly, $\mathbf{P}^{(i)}_{\rm{r}}$ and $\mathbf{U}^{(i)}$ can be calculated based on $\gamma^{(i)}(\nu,\mu)$ according to (\ref{inverse}) and (\ref{u3}). Further denote the maximum achievable data rate in the $i$-th iteration as $r^{(i)}$. At the beginning, we initialize $\gamma(\nu, \mu)$ as $\gamma^{(0)}(\nu, \mu)$. How to do this is discussed in (\ref{p16}) at the end of this subsection. In the $i$-th iteration,  the optimization of $\mathbf{P}_{\rm{c}}$ is formulated as
\begin{equation}\label{p12} 
\begin{array}{rcl}
  \mathcal{P}12: \!\!\!  & \!\!\! \max\limits_{\mathbf{P}_{\rm{c}}} &\!\!\!\!\! \sum\limits_{m=0}^{M\!-\!1} \!\sum\limits_{k=0}^{N_{\rm{c}} \!-\!1}\log\Big(\!1\! + \!\!\frac{(\!1\! -\! U^{(i-1)}(m,k)\!)P_{\rm{c}}(m,k)|\hat{H}_{\rm{c}}(m,k)|^2}{\sigma_{\rm{c}}^2}\!\Big) , \\ 
  & \text{s.t.} & \!\!\!\!\sum\limits_{m=0}^{M\!-\!1} \!\sum\limits_{k=0}^{N_{\rm{c}}\! -\!1}P_{\rm{c}}(m,k) = \bar{P}_{\rm{c}}, ~ P_{\rm{c}}(m,k)\geq 0.
\end{array}
\end{equation}
On the other hand, the optimization of the irrelevant cells of $\gamma( \nu,\mu)$ can be expressed as
\begin{equation}\label{p13} 
\begin{array}{rcl}
  \mathcal{P}13:\!\!\!  &\!\!\! \max\limits_{\gamma (\nu,\mu)} &\!\!\!\!\! \sum\limits_{m=0}^{M\!-\!1} \! \sum\limits_{k=0}^{N_c\! -\!1}\log\Big(\!1 \!+\! \frac{(\!1\!-\!U (\!m,k\!)\!)P^{(i)}_{\rm{c}}(m,k)|\hat{H}_{\rm{c}}(m,k)|^2}{\sigma_{\rm{c}}^2}\!\Big) , \\ 
  & \text{s.t.} & P_{\rm{r}}(m,k) \geq 0.\\
\end{array}
\end{equation}
The sub-problem $\mathcal{P}12$ can be solved by employing the KKT conditions, and the optimal solution is given by 
\begin{equation}\label{solution_sensing} 
  P^{(i)}_{\rm{c}}(m,k)\! \!= \! \left\{ \begin{array}{cl}
    \!\!\!\max \big\{\frac{1}{\bar{\beta} \ln2}\! - \!\frac{\sigma_{\rm{c}}^2}{|\hat{H}_{\rm{c}}(\!m,k\!)|^2}, 0 \big\}, & \!\!\!\!U^{(\!i\!-\!1\!)}(\!m,k\!) \!=\! 0, \\
   \! \!\!\!\!0 , & \!\!\!\!U^{(\!i\!-\!1\!)}(\!m,k\!)\! =\! 1 ,
  \end{array}\right. 
\end{equation}
where $\frac{1}{\bar{\beta} \ln2} = \frac{\bar{P}_{\rm{c}} + \sum_{(m,k)\in {\cal N}_{\rm e}}\frac{\sigma_{\rm{c}}^2}{|\hat{H}_{\rm{c}}(m,k)|^2} }{\text{card}({\cal N}_{\rm{e}})}$ and ${\cal N}_{\rm e}$ contains the indices of all the REs with non-zero power, i.e., $P_{\rm{c}}^{(i)}(m,k)>0$ \cite{ref8}. By contrast, the solution of the sub-problem $\mathcal{P}13$ is challenging to obtain, because the mapping function between $U(m,k)$ and $P_{\rm{r}}(m,k)$ is discontinuous and non-convex. To tackle this issue, an intuitive approach is to employ a linear continuous function of $P_{\rm{r}}(m,k)$ to approximate $U(m,k)$, which can be expressed as $U(m,k) \approx  P_{\rm{r}}(m,k)/A$, where $A$ is a normalization factor of $P_{\rm{r}}(m,k)$ and can be set as the largest element of $\mathbf{P}^{(0)}_{\rm{r}}$. Obviously, this is a rough approximation. But by carefully selecting the range of values for $P_{\rm{r}}(m,k)$, it can be made a tight approximation. More specifically, if the value of $P_{\rm{r}}(m,k)$ is close to $0$ or $A$, the error of the linear approximation is negligible. Considering $0 \leq P_{\rm{r}}(m,k) \leq A$, the term $P_{\rm{r}}(m,k)(1 - P_{\rm{r}}(m,k)/A)$ can be used to indicate how close $P_{\rm{r}}(m,k)$ is to 0 or $A$. If $P_{\rm{r}}(m,k)$ is equal to $0$ or $A$, the term is equivalent to $0$. On the other hand, if $P_{\rm{r}}(m,k)$ is significantly different from both $0$ and $A$, the value of the term becomes large. Therefore, $-P_{\rm{r}}(m,k)(1 - P_{\rm{r}}(m,k)/A)$ is added to the optimization objective of $\mathcal{P}13$ as a penalty term, which then becomes: 
\begin{equation}\label{p14} 
\begin{array}{rcl}
  \mathcal{P}14:\!\!\!  &\!\!\! \max\limits_{\gamma(\nu,\mu)} & \!\!\!\!\sum\limits_{m=0}^{M\!-\!1} \sum\limits_{k=0}^{N_c \!-\!1}\log\Big(\!1\! + \!\frac{(\!1\!-\!\frac{P_{\rm{r}}(m,k)}{A})P^{(i)}_{\rm{c}}(m,k)|\hat{H}_{\rm{c}}(m,k)|^2}{\sigma_{\rm{c}}^2}\!\Big)  \\
	& &	\!\!\! - \lambda P_{\rm{r}}(m,k)\Big(1\!-\!\frac{P_{\rm{r}}(m,k)}{A}\Big) ,\\ 
  & \text{s.t.} & \!\!\! 0 \leq P_{\rm{r}}(m,k) \leq A, \\  
\end{array}
\end{equation}
where $\lambda$ is a weight factor that trades off between the achievable data rate and the requirement for the range of values for $P_{\rm{r}}(m,k)$. 

Since the optimization objective of $\mathcal{P}14$ is not a concave function, $\mathcal{P}14$ is difficult to maximize directly. To tackle this issue, a lower bound of the optimization objective is derived by converting the second term $P_{\rm{r}}(m,k)(1 - P_{\rm{r}}(m,k)/A)$ with its linear approximation. According to the Minorize-Maximization (MM) algorithm, by maximizing the lower bound in each iteration, the results will finally converge to the optimal solution of the original problem \cite{mm}. Therefore, we maximize the lower bound of the optimization objective iteratively, where the problem in the $j$-th iteration can be formulated as
\begin{equation}\label{p15} 
\begin{array}{rcl}
  \mathcal{P}15:\!\!\!\! &\!\!\!\! \max\limits_{\gamma (\nu,\mu)} &\!\!\!\! \sum\limits_{m=0}^{M\!-\!1} \sum\limits_{k=0}^{N_c \!-\!1}\log\Big(\!1 \!+ \!\frac{(\!1\!-\!\frac{P _{\rm{r}}(\!m,k\!)}{A})P^{(i)}_{\rm{c}}(\!m,k\!)|\hat{H}_{\rm{c}}(\!m,k\!)|^2}{\sigma_{\rm{c}}^2}\Big) \\
  & &\!\!\!\! -\! \lambda \Big(\!P_{\rm{r}}(\!m,k\!) \!-\!\frac{2P^{(i,j\!-\!1)}_{\rm{r}}(\!m,k\!)}{A}P_{\rm{r}}(m,k) \!\Big),\\ 
  &\text{s.t.} & \!\!\!\!0 \leq P _{\rm{r}}(m,k) \leq A.\\  
\end{array}
\end{equation}
$P^{(i,j-1)}_{\rm{r}}(m,k)$ is calculated based on the optimization results $\gamma^{(i,j-1)}(\nu,\mu)$ in the $(j-1)$-th iteration. 
Besides, we denote the optimization result of the objective function in the $(j-1)$-th iteration as $\bar{r}^{(i,j-1)}$.
If $j$ is larger than the maximum number of iteration $J_{\rm{m}}$, or $\big|\bar{r}^{(i,j)}-\bar{r}^{(i,j-1)}\big|$ is less than a predefined threshold $\epsilon_{2}$, the iteration procedure is terminated, and the value of $\gamma^{(i)}(\nu,\mu)$ is obtained as $\gamma^{(i,j)}(\nu,\mu)$. Given the non-convex nature of problem $\mathcal{P}11$, the convergence of the alternating iterative algorithm depends on the initial value. To strike a balance between computational complexity and the optimality of the solution, it is advisable to select an initial value that is not only easy to obtain but also reasonably close to the optimal solution. Following this philosophy, we derive the initial value $\gamma^{(0)}(\nu, \mu)$ by solving the following problem:
\begin{equation}\label{p16} 
\begin{array}{rcl}
  \mathcal{P}16: & \min\limits_{\gamma( \nu,\mu)} & \sum\limits_{m=0}^{M-1} \sum\limits_{k=0}^{N_c -1}   P_{\rm{r}}(m,k)|\hat{H}_{\rm{c}}(m,k)|^2 , \\
  & \text{s.t.} & P_{\rm{r}}(m,k) \geq 0,  
\end{array}
\end{equation}
which is a linear programming problem, and can be readily solved with low complexity. The objective function aims to allocate low sensing power to REs with high communication channel gains, i.e., $|\hat{H}_{\rm{c}}(m,k)|^2$ is large. This avoids the sensing subsystem to occupy high-quality communication channels, while ensuring the optimal sensing performance. 

The complete procedure of sensing-centric waveform design is given in Algorithm~{\ref{alg2}}. The main computation complexity of Algorithm~{\ref{alg2}} lies in the two iterations corresponding to Line $2$ and Line $4$, and the total iteration number is no more than $I_{\rm{m}}J_{\rm{m}}$. In our simulation, when $\epsilon_1=10^{-3}$ and $\epsilon_2=10^{-1}$, the total iteration number is less than $20$ with the probability of $90\%$.

\begin{algorithm}[!t]
\caption{Sensing-centric waveform design procedure}
\label{alg2}
\begin{algorithmic}[1]
	\REQUIRE  Total communication power $\bar{P}_{\rm{c}}$, total sensing power $\bar{P}_{\rm{r}}$, estimated communication channel $\mathbf{\hat{H}}_{\rm{c}}$, region of interest $\Omega_{\bm{s}}$, maximum number of alternating iterations $I_{\rm{m}}$, maximum number of inner iterations $J_{\rm{m}}$, termination thresholds for outer and inner loops $\epsilon_{1}$ and $\epsilon_{2}$;
  \STATE Initialize $\gamma^{(0)}(\nu,\mu)$, $\mathbf{P}^{(0)}_{\rm{r}}$ and $\mathbf{U}^{(0)}$ according to $\mathcal{P}16$, (\ref{inverse}) and (\ref{u3});
  \WHILE{$i \leq I_{\rm{m}}$ or $ |r^{(i)}-r^{(i-1)}|<\epsilon_{1}$}
		\STATE Allocate the communication power $\mathbf{P}^{(i)}_{\rm{c}} $ according to $\mathcal{P}12$;
		\WHILE{$j \leq J_{\rm{m}}$ or $|\bar{r}^{(i,j)} -\bar{r}^{(i,j-1)}|<\epsilon_{2}$}
      \STATE Calculate $\gamma^{(i,j)}(\mathbf{P}_{\rm{r}},\nu,\mu)$, $\mathbf{P}^{(i,j)}_{\rm{r}}$ according to $\mathcal{P}14$;
      \STATE $j=j+1$;
    \ENDWHILE
    \STATE Calculate $\mathbf{U}^{(i)}$ according to (\ref{u3});
    \STATE $i=i+1$;
    \ENDWHILE
  \RETURN $\mathbf{P}_{\rm{c}}$, $\mathbf{P}_{\rm{r}}$, $\mathbf{U}$.	
\end{algorithmic}
\end{algorithm}

\section{Numerical Results}\label{S5}

Numerical results are provided to validate the proposed communication-centric and sensing-centric waveform designs and to provide useful guidelines for the implementation of the proposed designs. The simulation system parameters are listed in Table~{\ref{t1}}.

\begin{figure}[!b] 
\center
\includegraphics[width=0.9\linewidth, keepaspectratio]{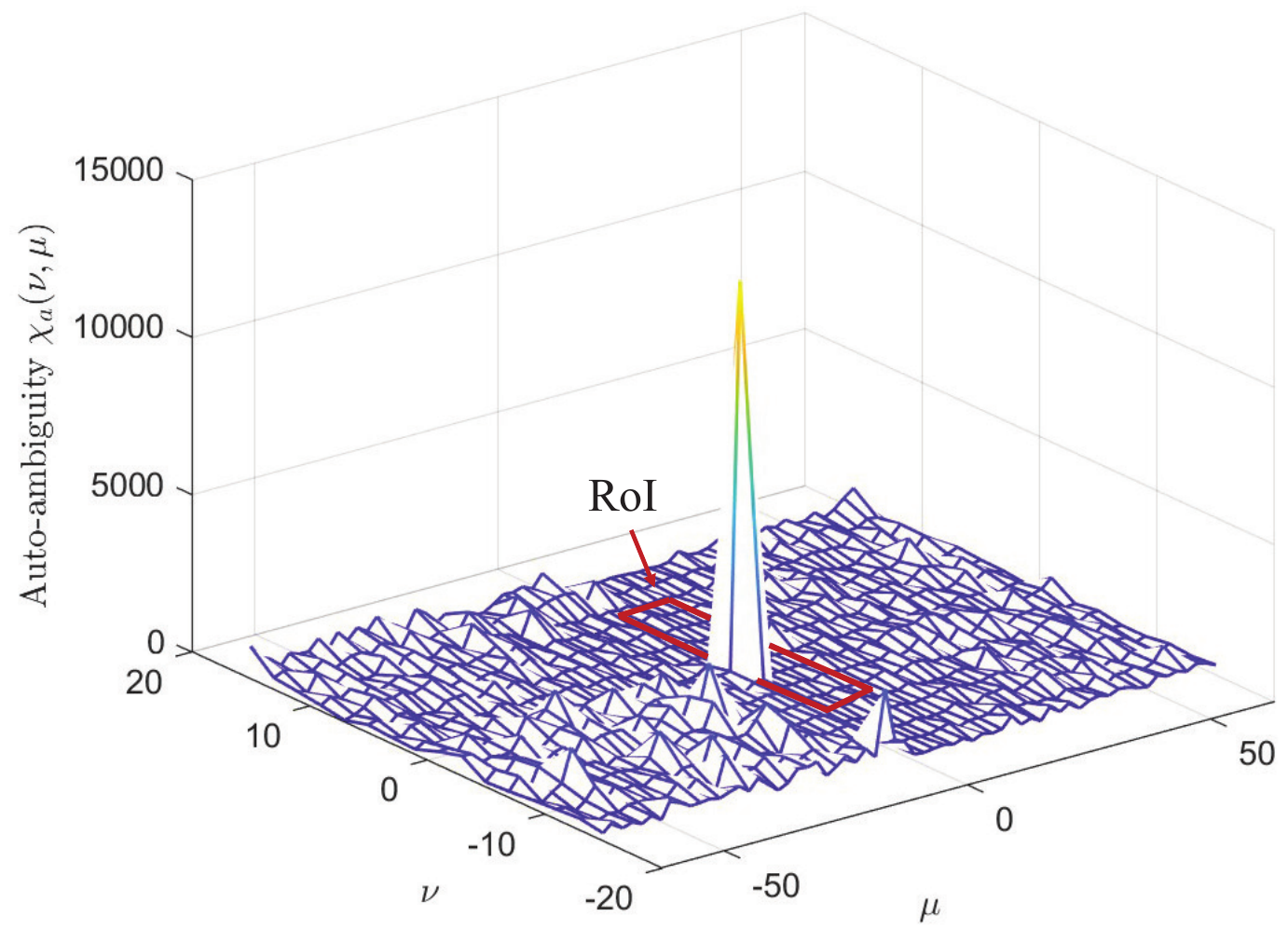}	
\vspace{-4mm}
\caption{Ambiguity function of the sensing component generated by the proposed communication-centric waveform design.}
\label{fig_radar} 
\vspace{-1mm}
\end{figure}

\subsection{Communication-Centric Design}\label{S5.1}

Fig.~\ref{fig_radar} presents the ambiguity function of the sensing component generated by our proposed communication-centric waveform design. The sensing scopes of distance and speed are set as $[0, 60]$\,m and $[-20,20]$\,m/s, respectively. The corresponding RoI is derived according to ($\ref{rof}$), which is outlined with red lines in Fig.~\ref{fig_radar}. It can be seen that since there are non-negligible sidelobes outside the RoI, the PSLR in the whole ambiguity function is $7$\,dB. However, the level of sidelobes within the RoI is marginal, leading to a PSLR of $12$\,dB within the RoI. 
Such a high PSLR provides a lower false alarm rate because the probability of noise or clutters being mistaken as targets is reduced \cite{zf}.
Besides, a high PSLR is advantageous for improving the resistance to interference and the resolution of  closely spaced targets \cite{Metric1, Metric2}.
Therefore, the proposed communication-centric waveform can guarantee superior sensing and positioning performance within the RoI, whilst attaining the maximum data throughput at the same time.

\begin{table}[!t]
\renewcommand{\arraystretch}{1.1}
\caption{System Parameters}
\vspace*{-5mm}
\begin{center}
\begin{tabular}{|c | c|  c | }
					\hline
					Symbol & Parameter & Value\\\hline
					$f_c$ & Carrier frequency & $240$ GHz\\\hline
                    $\Delta f$ & Subcarrier spacing & $240$ kHz \\\hline
                    $N_c$ & Number of subcarrier & $128/512$ \\\hline
                    $T$ & OFDM symbol duration & $4.1470$ $\mu$s\\\hline
                    $T_G$ & Cyclic prefix length & $1.0368$ $\mu$s\\\hline
                    $T_O$ & Total OFDM symbol duration & $5.1838$ $\mu$s\\\hline
                    $M$ & Number of OFDM symbol & $32$\\\hline
                    $d_m$ & Maximum range & $155$ m \\\hline
                    $v_m$ & Maximum relative speed & $\pm60$ m/s \\\hline
\end{tabular} 
\end{center}
\vspace*{-6mm}
\label{t1}
\end{table}

\begin{figure}[!t]
\center
 \subfigure[Fast time-varying channel]{\label{fig01}
	\includegraphics[width=0.87\linewidth, keepaspectratio]{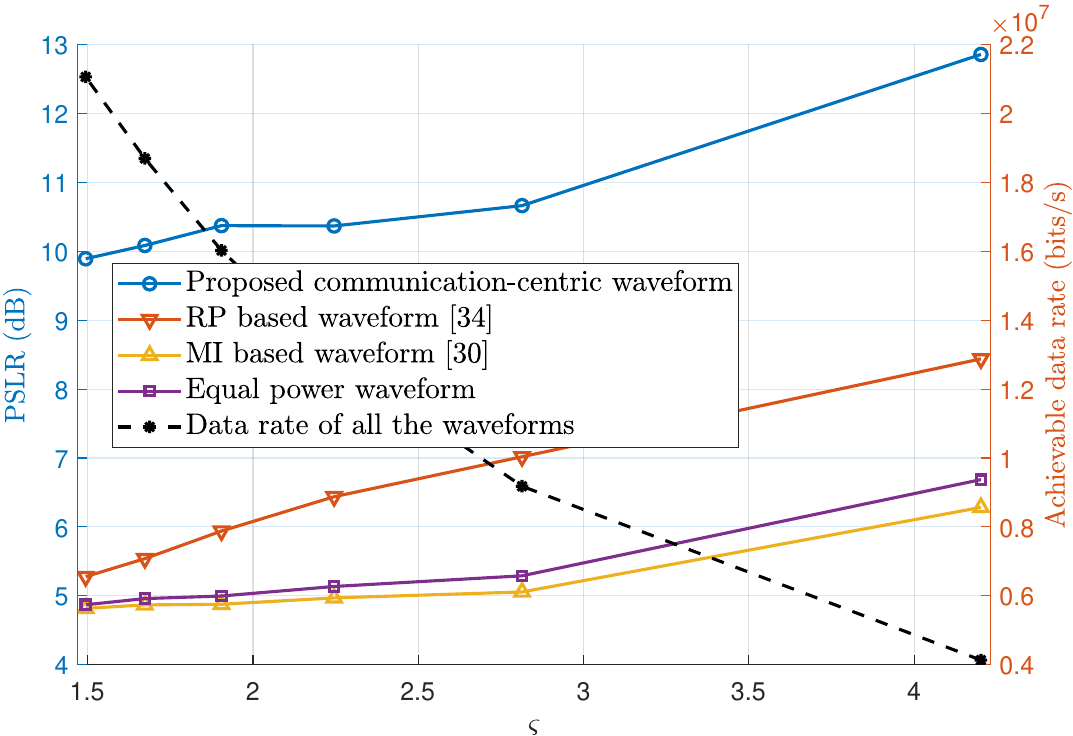}	
 }
 \hspace*{-1mm}\subfigure[Slow time-varying channel]{\label{fig02}
	\includegraphics[width=0.87\linewidth, keepaspectratio]{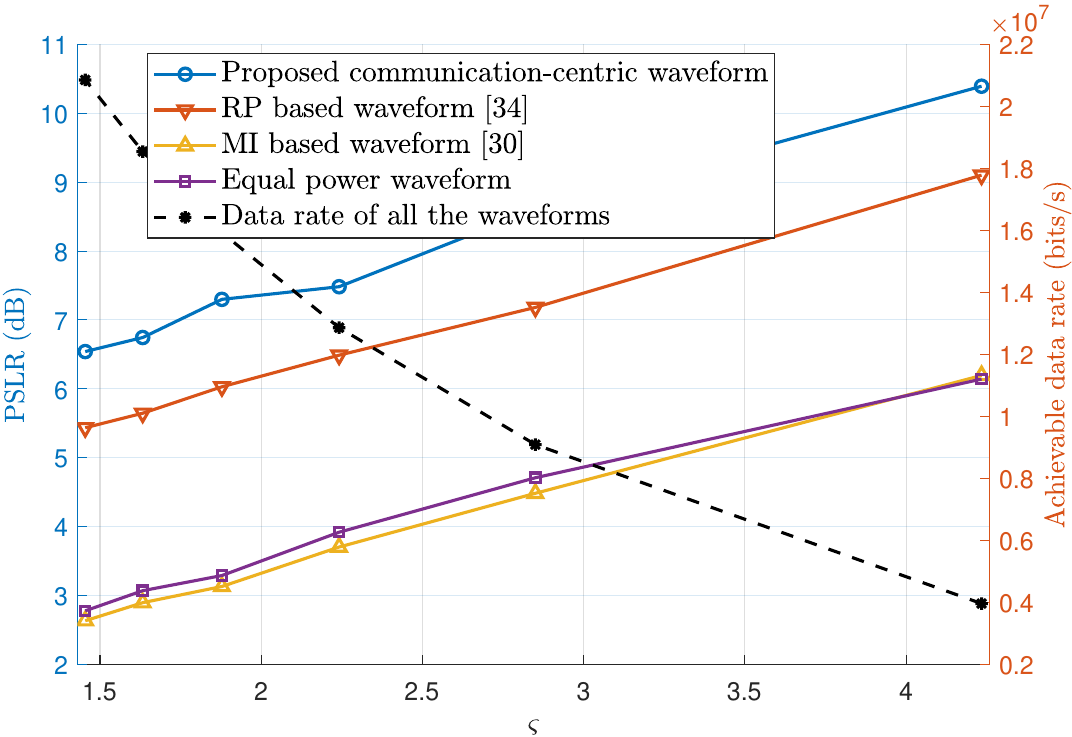}
 }
\vspace*{-1mm}
\caption{Performance comparison between the proposed communication-centric waveform and existing designs in terms of PSLR as the function of the channel quality threshold $\varsigma$ with $128$ subcarriers.}
\label{fig5}  
\vspace*{-4mm}
\end{figure}

Fig.~\ref{fig5} illustrates the sensing and communication performance of the proposed communication-centric
waveform and three existing designs, in terms of the PSLR within the RoI and the achievable data rate, where a wide range of the channel quality threshold $\varsigma$ is adopted and the number of subcarrier is set to $128$. The fast and slow time-varying channels are considered in Figs.~\ref{fig01} and \ref{fig02}, respectively, with the former having a larger Doppler frequency offset range ($\sim 100$ kHz) than the latter ($\sim 1$ kHz). The three existing designs compared are the range profile (RP) based waveform \cite{Chen_cl_23}, the MI based waveform \cite{Bica_conf_19}, and the equal power waveform. Specifically, the RP based waveform allocates the sensing power to each OFDM symbol separately and optimizes the PSLR in the range profile, the MI based waveform optimizes the MI between the target impulse response and the received signal, while the equal power waveform evenly distributes the sensing power among the sensing REs.

For a fair comparison, all the four waveforms are based on the communication-centric criterion. In other words, for a given communication power $\bar{P}_{\rm{c}}$, all the four waveforms are designed by firstly allocating REs with high-quality channel conditions, i.e., $\big\|\hat{H}_{\rm{c}}(m,k)\big\|^2 \geq \varsigma$, for data transmission. Therefore, the achievable data rates of all the four waveforms are the same, which is referred to as `data rate of all the waveforms' and corresponds to the black dashed curve in Figs.~\ref{fig01} and \ref{fig02}. It is observed that as the threshold $\varsigma$ increases, fewer REs are allocated
to communications, and this results in a decrease in the achievable data rate.

By contrast, as $\varsigma$ increases, more REs are allocated to sensing, and this leads to an increase in the PSLR within the RoI. It can be seen that the proposed communication-centric waveform is capable of significantly improving the PSLR compared with the three benchmark waveforms in both fast and slow fading channels. 
In the slow fading channel, where the channel condition for each OFDM symbol remains nearly unchanged within a frame, our proposed waveform improves the PSLR by about $1$\,dB compared to the RP based waveform. Conversely, in the fast fading channel, where the channel conditions among OFDM symbols vary quickly, the joint design over multiple OFDM symbols is necessary for sensing performance enhancement. In this case, our proposed waveform improves the PSLR by nearly $5$\,dB over the second best RP based waveform. 

\begin{figure}[!t]
 \subfigure[Fast time-varying channel]{\label{fig01}
	\includegraphics[width=0.87\linewidth, keepaspectratio]{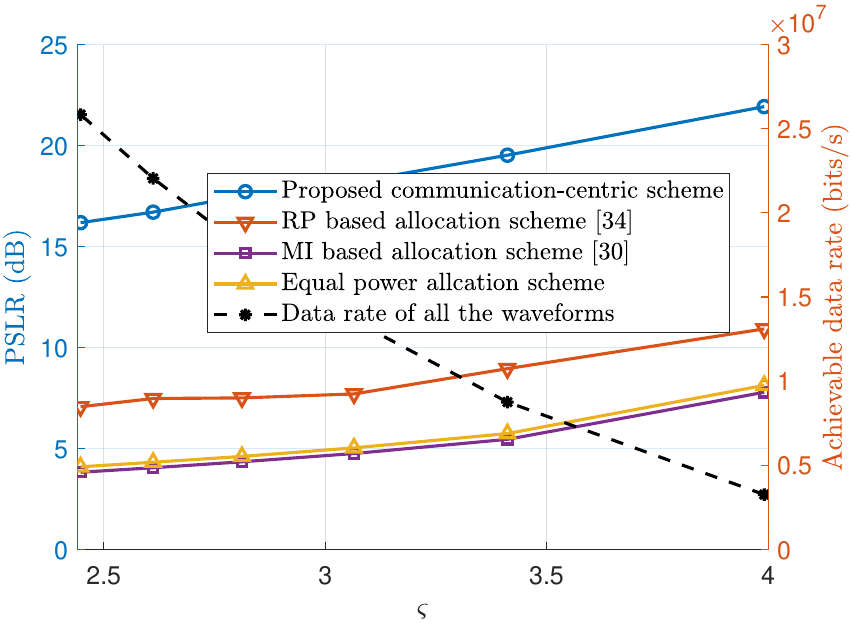}	
 }
 \hspace*{-1mm}\subfigure[Slow time-varying channel]{\label{fig02}
	\includegraphics[width=0.87\linewidth, keepaspectratio]{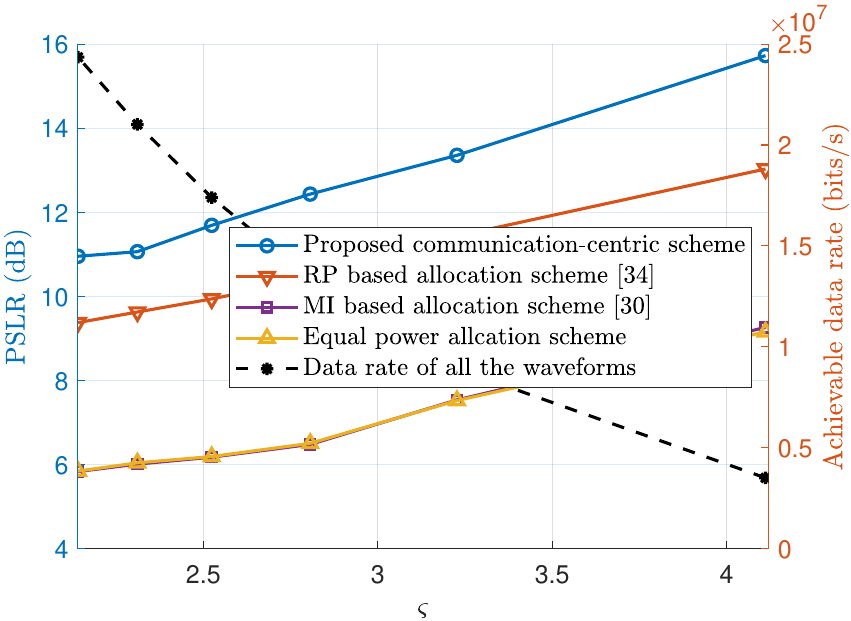}
 }
\vspace*{-1mm}
\caption{Performance comparison between the proposed communication-centric waveform and existing designs in terms of PSLR as the function of the channel quality threshold $\varsigma$ with $512$ subcarriers.}
\label{fig51} 
\vspace*{-6mm}
\end{figure}

By comparing Fig.~\ref{fig01} with Fig.~\ref{fig02}, it can be seen that the PSLR performance of the three existing designs in the fast fading channel situation are generally worst than their PSLR performance in the slow fading channel scenario, which is to be expected. However, the PSLR performance of our proposed design is actually better in the fast fading case than in the slow fading one. The reason for this `unexpected' phenomenon can be explained as follows. In the slow fading channel, due to the relatively stable channel conditions, certain subcarriers under high-quality channel may remain allocated for communications within multiple consecutive OFDM symbols. Accordingly, the sensing subsystem is unable to employ these subcarriers throughout the whole processing interval, which induces high sidelobes in the ambiguity function. By contrast, in the fast fading channel, the RE allocation between communication and sensing subsystems changes dynamically across multiple OFDM symbols, allowing the sensing subsystem to utilize diverse subcarriers at different instants in the processing interval. Then through our proposed joint multi-symbol power optimization, the sensing system is capable of effectively integrating information from different subcarriers, resulting in an increase in the PSLR within the RoI, compared with the slow fading case.
Furthermore, through the proposed PAPR reduction method of Algorithm~\ref{alg1}, our communication-centric waveform manages to obtain acceptable PAPR levels of $5.83$\,dB (when employing binary PSK) and $4.81$\,dB (when employing quadrature PSK), which achieves more than $8$\,dB PAPR reduction over those without phase adjustment. 
This reduces the impacts of possible nonlinearity from imperfect hardware devices, e.g., high power amplifier, on the sensing/positioning performances.

Fig.~\ref{fig51} investigates the sensing and communication performance of the proposed communication-centric waveform and the three existing designs with $512$ subcarriers. Compared with Fig.~\ref{fig5}, the bandwidth is enlarged by increasing the number of subcarriers.
It can be seen that our proposed scheme still outperforms three existing designs in term of PSLR with RoI. 
Some phenomena depicted in Fig.~\ref{fig51} bear a basic resemblance to those observed in the small bandwidth case in Fig.~\ref{fig5}, which can be explained by using the same underlying principles.
It is worth noting that the PSLR of our proposed scheme exhibits an improvement of approximately $6$ dB, compared with the small bandwidth case. 
This is because, as the number of subcarriers increases while communication power remains constant, the sensing function is able to occupy a greater number of REs, leading to an increase in the PSLR within RoI.
 
\vspace{1mm}
 \begin{figure}[!t]
	\center
 \subfigure[Achievable data rate versus sensing range of distance]{\label{fig61}
	\includegraphics[width=0.85\linewidth, keepaspectratio]{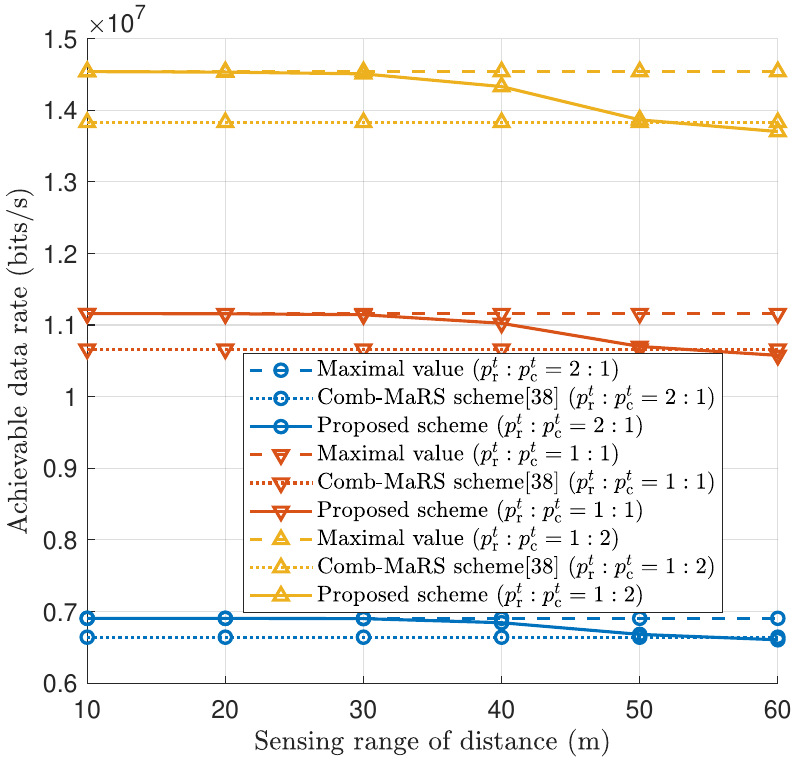}	
   }
  \subfigure[Achievable data rate versus sensing range of speed]{\label{fig62}
	\includegraphics[width=0.85\linewidth, keepaspectratio]{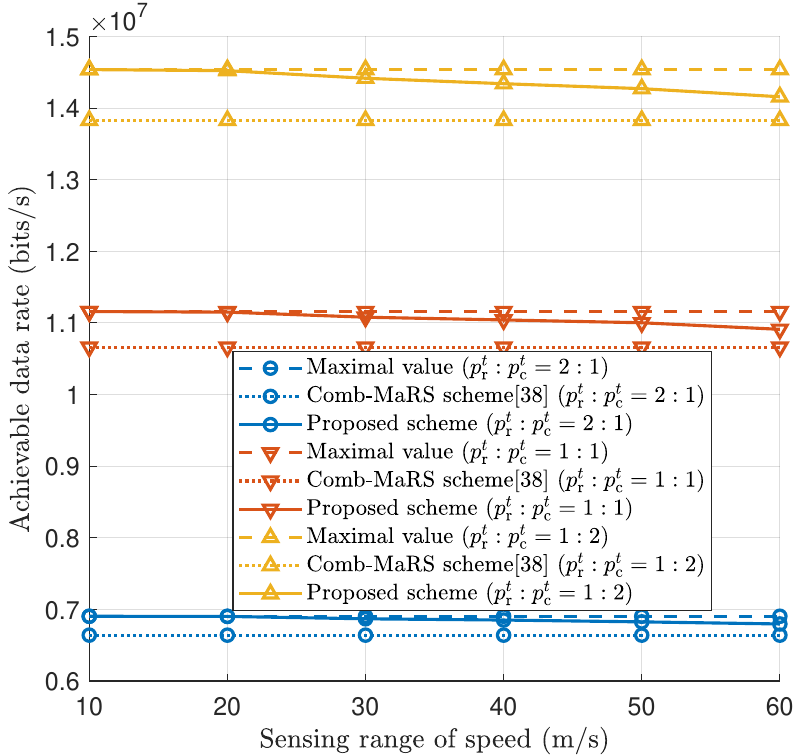}
}
\vspace*{-1mm}
\caption{Achievable data rate comparison with respect to the sensing ranges of distance and speed. The sensing range of speed is set as $[-60, 60]$\,m/s in (a) and the sensing range of distance is set as $[0,60]$\,m in (b).
\label{fig6}}   
\vspace*{-5mm}
\end{figure} 
 
\subsection{Sensing-Centric Design}\label{S5.2}

Figs.~\ref{fig61} and \ref{fig62} compare the achievable data rates of the proposed sensing-centric waveform with those of the two baselines with respect to the sensing ranges of distance and speed, respectively, where different ratios of the sensing power to the communication power are considered. 
More specifically, the baseline, referred to as `maximal value', is set to an upper bound of the achievable data rate obtained by maximizing the communication performance without the consideration of sensing. 
The other baseline is a comb-shaped max-aperture radar slicing (Comb-MaRS) waveform proposed in \cite{Ma_iotj_23}.
As expected, the achievable data rates of our proposed waveform and the baselines increase with the decrease in the ratio of the sensing power to the communication power, since more power is allocated to the communication subsystem. 
In the Comb-MaRS waveform, the RE assignment strategy remains constant regardless of variations in the sensing ranges. 
Consequently, the communication data rate of the Comb-MaRS waveform does not fluctuate as the sensing range of distance/speed increases. 
By contrast, our proposed scheme adapts the RE assignment strategy in accordance with variations in sensing range.
As the sensing ranges of distance and speed increase, i.e., the RoI becomes larger, the number of irrelevant cells decreases, which results in the reduction of the dimension of optimization variables for communication performance. Therefore, the achievable data rate of our proposed waveform degrades slightly with the increase in sensing ranges of distance and speed. However, the achieved data rate of our sensing-centric waveform still closely approaches the maximal value  and outperforms that of the Comb-MaRS scheme when the sensing ranges of distance and speed are no more than 40\,m and $\pm50$\,m/s, respectively, which validates the feasibility and effectiveness of Algorithm~\ref{alg2}.
Moreover, regarding sensing performance, the simulations indicate that the PSLR of our scheme ($\sim 20$ dB) significantly surpasses that of the Comb-MaRS scheme ($\sim 3$ dB).

\begin{figure}[!t]
\center
\includegraphics[width=1\linewidth, keepaspectratio]{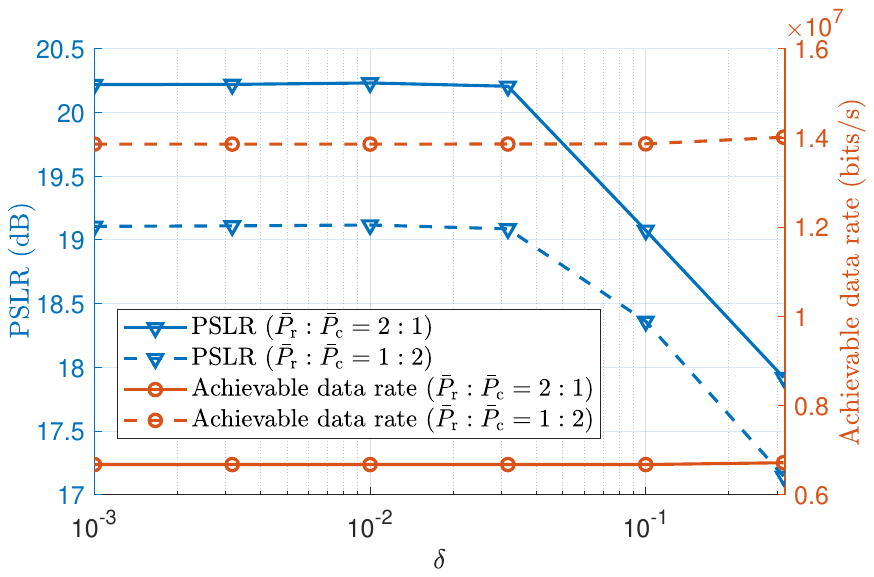}	
\vspace{-6mm}
\caption{PSLR and achievable data rate as the functions of threshold $\delta$, where $\lambda = 0.02$ and the sensing ranges of distance and speed are set as $[0,50]$ m and $[-60, 60]$ m/s, respectively.}
\label{fig_delta} 
 \vspace{-4mm}
\end{figure}

Fig.~\ref{fig_delta} investigates the sensing and communication performances of the proposed sensing-centric waveform, in terms of the PSLR within the RoI and the achievable data rate, with respect to the sensing power threshold $\delta$ for each RE. Specifically, the PSLRs within the RoI under different ratios of the sensing power to the communication power are illustrated by the blue curves, while the achievable data rates are illustrated by the orange curves. It can be seen that the PSLR degrades with the increase of $\delta$ when $\delta$ is larger than $0.03$. This is because more REs are designated for the communication subsystem, even though some of them have significant impact on the sensing performance. On the other hand, when $\delta$ is between $10^{-3}$ to $10^{-1}$, the achievable data rate exhibits only a very marginal improvement as $\delta$ increases. This is attributed to the fact that the additional communication REs may operate under unfavorable channel conditions. Consequently, we can set $\delta$ between $10^{-3}$ and 0.03 in order to achieve a good trade-off in the RE assignment that guarantees the accurate sensing and positioning with minimal impact on communication performance.

\begin{figure}[!b]
\vspace*{-6mm}
\center
\includegraphics[width=1\linewidth, keepaspectratio]{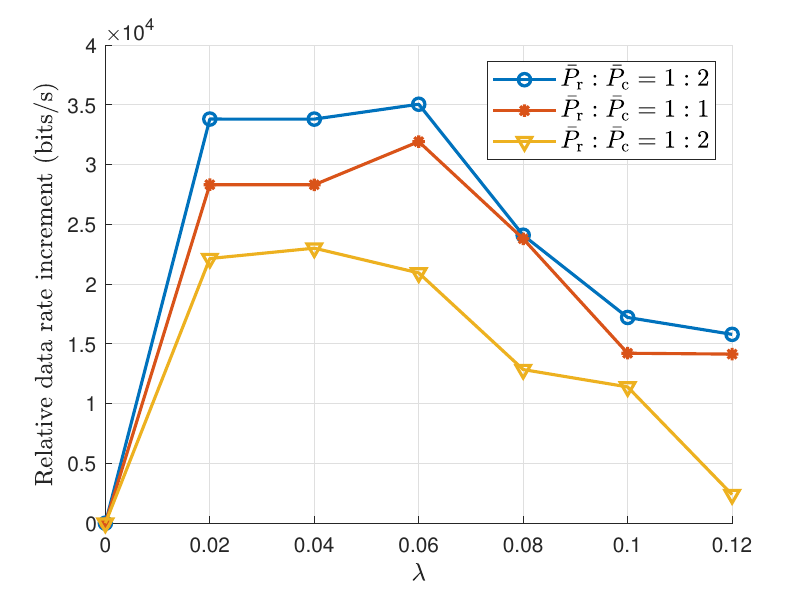}	
\vspace{-4mm}
\caption{Relative data rate increment with respect to $\lambda$, where $\delta = 0.03$ and the sensing ranges of distance and speed are set as $[0,50]$ m and $[-60, 60]$ m/s, respectively.}
 \label{fig_lambda} 
\end{figure}

Fig.~\ref{fig_lambda} presents the communication performance of our proposed sensing-centric waveform, in
terms of the relative data rate increment with respect to $\lambda$, which is the weighting factor of the penalty term in $\mathcal{P}14$. The relative data rate increment is calculated as the difference between the current achievable data rate and the data rate at $\lambda=0$. As $\lambda$ increases, we observe a significant rise in the achievable data rate when $\lambda$ is less than $0.02$, followed by a rapid decline when $\lambda>0.06$. This is because when $\lambda <0.02$, the penalty term cannot effectively constrain the values of $P_{\rm r}(m,k)$, reducing $\lambda$ leads to a larger approximation error that negatively impacts the achievable data rate. On the other hand, when $\lambda >0.06$, the penalty term outweighs the data rate maximization, and increasing $\lambda$ degrades the communication performance. Therefore, in this example, $\lambda$ can be set between $0.02$ and $0.06$ numerically to balance the penalty term and communication performance.

\section{Conclusions}\label{S6}

In this paper, a cross-domain OFDM-based waveform design methodology, including communication- and sensing-centric waveform designs, was proposed for sensing performance enhancement with minimal impact on communication capabilities.
In the communication-centric waveform design, 
the sensing performance, characterized by the PSLR and PAPR of the sensing component, was optimized with well-designed power- and phase-domain waveform coefficients.  
On the other hand, the sensing-centric design ensured a `locally' perfect auto-correlation property for the sensing sequence by adjusting the ambiguity function value within the RoI of the integrated waveform
while minimizing its impact on communication.
Numerical results showed that the sensing component of the proposed communication-centric waveform achieves significantly higher PSLR and low PAPR, compared with the existing waveform designs. Moreover, the proposed sensing-centric waveform approaches the maximum achievable data rate while providing a `locally' perfect auto-correlation property for accurate sensing and positioning.

There exist several open issues for our proposed cross-domain waveform design methodology, which are set aside as our future work due to the page limit.
The implementation of the proposed waveform relies on consistent channel estimations at both the BS and UE sides. However, there may be errors in the prediction and estimation of the communication channel in practical systems, leading to performance degradation.
We will carry out performance analysis and investigate improved schemes against estimation errors.
Secondly, since the sensing-centric waveform design is formulated as a non-convex optimization problem, a quasi-optimal solution is derived by the proposed alternating optimization algorithm. To obtain the optimal waveform design, more advanced optimization algorithms (e.g., the Majorize-Minimize algorithm and second-order optimization) will be explored. 
Finally, the antenna array with directional beamforming cannot be regarded as a single directional antenna under some non-ideal factors (e.g., beam squint effect). 
Therefore, the joint design of the hybrid beamforming and the integrated waveform will be investigated in depth in our future work.

\bibliographystyle{IEEEtran}


\end{document}